\documentclass[twocolumn,mathlines]{aastex631}

\usepackage{amsmath}

\begin{document}

\title{Driven phase-mixed Alfv\'en waves in a partially ionized solar plasma}

\author[0009-0001-1997-4361]{M. McMurdo}
\correspondingauthor{M. McMurdo}
\affiliation{Centre for mathematical Plasma Astrophysics, Department of Mathematics, KU Leuven, Celestijnenlaan 200B Bus 2400, 3001 Leuven, Belgium}
\email{max.mcmurdo@kuleuven.be}
\author[0000-0002-3066-7653]{I. Ballai}
\affiliation{Plasma Dynamics Group, School of Mathematical and Physical Sciences, The University of Sheffield, Hicks Building, Hounsfield Road, Sheffield, S3 7RH, UK}
\author[0000-0002-9546-2368]{G. Verth}
\affiliation{Plasma Dynamics Group, School of Mathematical and Physical Sciences, The University of Sheffield, Hicks Building, Hounsfield Road, Sheffield, S3 7RH, UK}
\author[0000-0002-0893-7346]{V. Fedun}
\affiliation{Plasma Dynamics Group, School of Electrical and Electronic Engineering, The University of Sheffield, Mappin Street, Sheffield, S1 3JD, UK}



\begin{abstract}

Phase mixing has long been understood to be a viable mechanism for expediting the dissipation of Alfv\'en wave energy resulting in the subsequent heating of the solar atmosphere. To fulfil the conditions necessary for phase mixing to occur, we consider the cross-field gradient in the Alfv\'en speed as a free parameter in our model. Using a single-fluid description of a partially ionized chromospheric plasma, we explore the efficiency of damping of shear Alfv\'en waves subject to phase mixing when a pulse wave driver is employed. Our results demonstrate a strong dependence of the dissipation length of shear Alfv\'en waves on both the ionization degree of the plasma and the gradient of the Alfv\'en speed. When assessing the efficiency of phase mixing across various inhomogeneities, our findings indicate that waves originating from a pulse driver exhibit initially identical heating rates as those generated by a continuous wave driver. One key difference observed was that Alfv\'en pulses possess a lower overall decay rate due to a change in damping profile from exponential to algebraic. This discrepancy arises from the absence of a consistent injection of energy into the base of the domain, that preserves longitudinal gradients of the magnetic field perturbations more effectively. These findings demonstrate the importance of understanding the relations between the wave driver, damping mechanisms, and propagation dynamics in resolving the atmospheric heating problem.

\end{abstract}



\section{Introduction}\label{sec:Introduction}


Magnetohydrodynamic (MHD) waves are prevalent throughout the solar atmosphere and are widely considered a key mechanism for sustaining the high temperatures of the upper layers. Although the solar corona is roughly two orders of magnitude hotter than the partially ionized chromosphere, it is often overlooked that due to the relatively large density of the chromosphere, the heating rate required to compensate for radiative losses is approximately two orders of magnitude higher \citep{Withbroe1977}. For this reason, it is important to understand how the chromosphere is heated. 

Magnetic reconnection is a fundamental process in solar physics, driving numerous solar events such as solar flares, coronal mass ejections, and the excitation of MHD waves \citep{Yang2016}. This process involves magnetic field lines with opposite polarities breaking and reconnecting, which converts magnetic energy into heat, often initiating various waves, including transverse MHD waves \citep{Sukarmadji2024}. In the lower solar atmosphere, particularly in the photosphere and chromosphere, magnetic reconnection is believed to be responsible for Ellerman bombs (EBs) that are characterized by intense brightening observed in the extended wings of the hydrogen Balmer-$\alpha$ line (H-$\alpha$) and typically occur within complex bipolar active regions during phases of significant flux emergence \citep[e.g., see][]{EBWatanabe,EBVissers}. Advanced simulations by \cite{EBHansteen1, EBHansteen2} and \cite{EBDanilovic} provide evidence that EBs are the result of small-scale magnetic reconnection at photospheric levels. Studies by \cite{EBRouppe} and \cite{EBJoshi} have documented EB-like brightenings, termed quiet Sun EBs (QSEBs), in the quiet Sun far from regions of intense magnetic activity. These QSEBs show weaker H-$\alpha$ wing enhancements compared to those in highly active regions \citep{EBNelson}. The transient nature of EBs and QSEBs, along with their connection to small-scale magnetic reconnection in the photosphere, suggests that they could be a source of pulse-like Alfv\'en waves in the lower solar atmosphere. For an extensive review of EBs, including diagnostics and modeling, see e.g., \cite{EBRutten,EBFang}. The transient character of these events and their link to localized magnetic reconnection in the partially ionized, inhomogeneous lower atmosphere may significantly enhance the damping of Alfv\'en waves through phase mixing \citep[as shown by][for continuously driven waves]{McMurdo1}. Examining the propagation and behavior of finitely driven Alfv\'en waves in a partially ionized plasma is, therefore, essential for understanding their role in energy transport and dissipation in the solar atmosphere, highlighting the need for detailed modeling and simulations of phase-mixed Alfv\'en waves under these conditions.

Phase mixing has been proposed as a mechanism to heat the solar corona as early as 1980s by \cite{HeyvaertsPriest1983}, and while many authors have continued their investigations of the effects of Alfv\'en wave damping due to phase mixing \citep[see, e.g.,][]{Hood1997, mocanu2008, EdabiHosseinpour2013, RudermanPetrukhin2018, Vandamme2020}, very little attention has been paid to the effects of partial ionization on phase mixing. In a recent study, \cite{McMurdo1} have shown that the damping rates and subsequent heating of Alfv\'en waves generated by a harmonic wave driver can provide sufficient energy to balance radiative losses in the chromosphere of the quiet Sun. In this study, we extend upon this previous research by examining the damping of shear Alfv\'en waves in partially ionized plasmas, specifically when driven by a disturbance with a finite lifetime. This study incorporates the presence of a variety of transport mechanisms, to demonstrate wave damping in the quiet Sun's lower atmosphere.

Recent observations have proved the existence of transverse oscillations in partially ionized plasmas \citep{Alfven2011Nature, ObservationBate}. It is typical to only see at most a few wavelengths, suggesting that wave drivers often operate for a small finite time, as well as the typically assumed continuous driver \citep{decaylesskinkoscillations, decaylesskinkoscillations2}. For this reason, we seek to understand the characteristic damping of Alfv\'en waves generated by more realistic wave drivers. 

When we consider a monochromatic harmonic driver we assume there is a constant injection of energy at the base of the photosphere. As a result, the trailing edge of the wave cannot dampen effectively and wave damping occurs predominantly at the leading edge of the wave. However, if we were to consider a pulse, both the leading edge and trailing edge are affected by phase mixing leading to a widening of the profile as shown by \cite{Hood2002} resulting in a larger effective wavelength of the pulse. 


Pulsating wave drivers have been employed extensively in numerical models conducting wave propagation and heating studies \citep[see, e.g.,][]{AWPulseThurgood,AWPulseChmielewski,AWPulseTsiklauri,AWPulseSrivastava,AWPulseKumar}. The problem of phase mixing of Alfv\'en pulses in fully ionized coronal holes has been investigated by \cite{Hood2002}, who found that the amplitude decay rate behaved algebraically rather than exponentially, as found earlier by \cite{HeyvaertsPriest1983}. \cite{Hood2002} assumed an initial profile disregarding the effects of the transversal Alfv\'en speed profile and the effects of diffusion during the excitation phase, eliminating the requirement of a time-dependent initial condition (emulating the wave driver at the base of the domain). In addition to this, the dissipative coefficients assumed here were roughly 6-7 orders of magnitude larger than formulae predicts, requiring the underlying assumption of turbulence to explain this enhancement. In this investigation, we seek to address these shortcomings by introducing a time-dependent pulse driver in addition to partial ionization effects. This is accomplished by implementing a methodology analogous to that of the continuous drivers presented by \cite{McMurdo1}, with the distinction that the driver is turned off after a specified duration has elapsed, achieving the finite-lifetime nature of this driver. A key distinction between this method and that employed by \cite{Hood2002} is that in our simulations, the perturbation is subject to phase mixing, diffusion and viscosity from the moment the magnetic field line is perturbed, providing more realistic simulations.

The structure of our study is as follows: in Sections \ref{sec:Governing Eqn} and \ref{sec:Alfven speed profiles} we set out the physical, mathematical and numerical formulation required to study phase-mixed Alfv\'en waves in partially ionized plasmas; with the results of our analysis being discussed in Section \ref{sec:Results}. Finally, our findings are summarized and concluded in Section \ref{sec:Discussion+Conclusions}.

\section{Physical considerations and the governing wave equation}\label{sec:Governing Eqn}

\begin{figure}[htp]
   \centering
    \includegraphics[width=.45\textwidth]{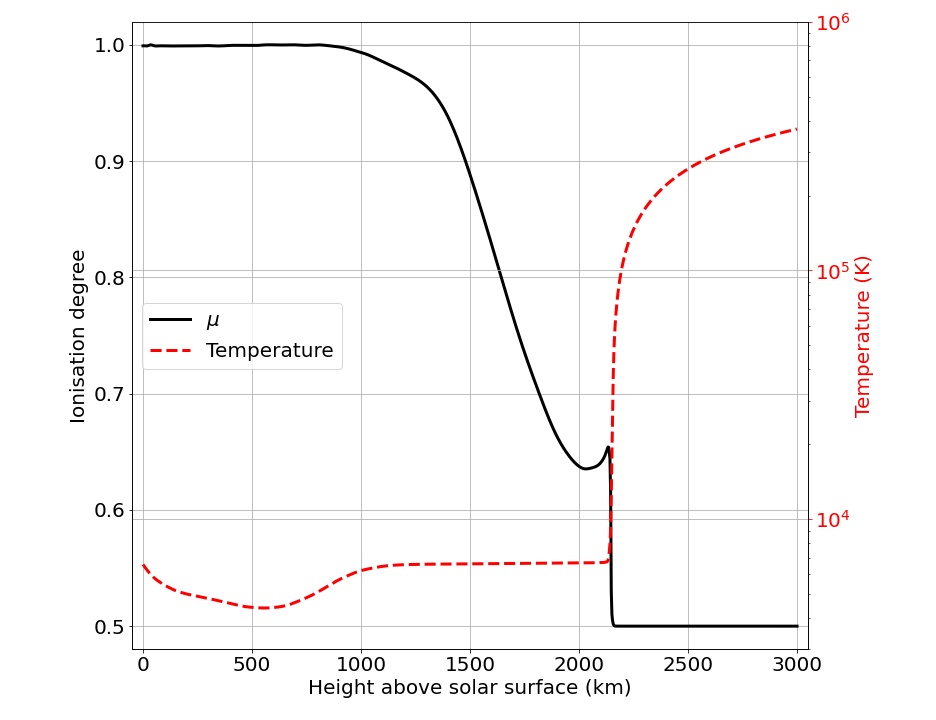}
    \caption{The ionization degree and plasma temperature are plotted with height according to the AL c7 model \citep{ALC7} for heights above the solar surface up to $3000$ km.}
  \label{fig:IonisationDegree_Temp}  
\end{figure}

In the solar chromosphere, the temperature is insufficient to achieve full ionization of the plasma, which we assume to be composed entirely of hydrogen. As a result, the plasma consists of a combination of protons, electrons, and neutral atoms, all of which interact via collisions. To characterize the ionization state of the plasma, we define the relative densities of ions and neutrals, represented by $\rho_i$ and $\rho_n$, respectively, as:

\begin{equation}
    \xi_{i} = \frac{\rho_{i}}{\rho} \approx \frac{n_{i}}{n_{i} + n_{n}}, \text{   } \xi_{n} = \frac{\rho_{n}}{\rho} \approx \frac{n_{n}}{n_{i}+n_{n}},    \label{eq:Relative densities}
\end{equation}
so that $\xi_i+\xi_n=1$ and $n_i$ and $n_n$ are the ion and neutral number densities, respectively. We now define the ionization degree of the plasma to be
$\mu = 1/(1 + \xi_{i})$, so that $\xi_i=(1/\mu)-1$ and $\xi_n=2-1/\mu$. 
The parameter $\mu$ can be varied between $1/2$ (fully ionized plasma) and $1$ (fully neutral fluid) and used as a key indicator of the efficiency of phase mixing in relation to the relative ion and neutral populations of the plasma. Furthermore, we define the quantity $\sigma$ as the ratio of the relative densities of ions and neutrals as
\[
\sigma=\frac{\xi_i}{\xi_n}=\frac{1-\mu}{2\mu-1}.
\]
The ionization degree of the plasma can be determined using standard atmospheric models of the Sun - in this study, we employ the AL c7 model \citep{ALC7}. The variations in ionization degree and temperature with height are shown in Figure \ref{fig:IonisationDegree_Temp}.

In a single fluid plasma, the effects of partial ionization are cast in the various transport coefficients including the Cowling diffusion, related to ambipolar diffusion, which appears thanks to the lack of response of neutral particles to the presence of the magnetic field. The various transport coefficients employed throughout the present investigation depend on a variety of parameters, namely the temperature of the plasma (assumed to be isothermal), collisional cross sections (assumed to be constant), magnetic field strength (also assumed constant), particle number densities (taken from the AL c7 model) and a variety of fundamental constants, used in plasma physics. The remaining quantities, such as the collisional frequencies, are derived from these values.

The various transport coefficients considered throughout this present investigation are given by, e.g., \cite{vranjes2014,McMurdo1} and consist of shear viscosity, ambipolar diffusion and magnetic diffusion. The particular form of the dissipative coefficients relevant to our investigation are given in detail earlier in the study by \cite{McMurdo1}, here we will summarize the key relations that will be used.

Shear viscosity quantifies the internal frictional forces between adjacent layers of fluid in relative motion and is given by 
\begin{equation}
    \nu_{v} = \frac{n_{n}k_BT\tau_{n}}{2}\frac{\Delta\gamma + (\omega_{ci}\tau_i)^2}{\Delta^2+(\omega_{ci}\tau_i)^2},
    \label{eq:shearviscosity}
\end{equation}
where $k_B$ is the Boltzmann constant, $T$ is the temperature, $\omega_{ci}=eB_0/m_i=v_A\left(e^2\mu_0n_i/m_i\right)^{1/2}$ is the ion cyclotron frequency, $e$ is the elementary charge, $B_0$ is the background magnetic field, $m_i$ is the ion mass, $v_A$ is the Alfv\'en speed, $\mu_0$ is the magnetic permeability, $\nu_{ab}$ denotes the collisional frequency between species $a$ and $b$. Their expressions are given by \cite{Braginskii1965,McMurdo1}, $\tau_{i,n}$ are collisional times where subscripts $i$ and $n$ correspond to ions and neutrals, respectively, and are given as a weighted sum of collisions between each fluid and itself \citep[see, e.g.,][for more details]{vranjes2014}. Finally, the quantities $\Delta$ and $\gamma$ are given by
\[
\begin{split}
    & \Delta=1-\frac{1}{(3\nu_{ii}/\nu_{in}+4)(3\nu_{nn}/\sigma\nu_{in}+4)}, \\ & \gamma=1+\frac{\sigma}{3\nu_{ii}/\nu_{in}+4}.
\end{split}
\]

Magnetic diffusivity (its coefficient is denoted by $\eta$), represents the resistance to motion of a plasma in the presence of a magnetic field resulting in the diffusion of a magnetic field through a plasma. It is dependent on the collisional frequencies of electrons with ions and neutrals. In this case, its value is given by 
\begin{equation}
    \eta = \frac{m_{e}(\nu_{ei}+\nu_{en})}{e^2n_{e}\mu_{0}}, \label{eq:Spitzer}
\end{equation}
where $m_e$ is the electron mass and $n_e$ is the electron number density. Finally, the ambipolar diffusion coefficient is given by
\begin{equation}
    \eta_{A} = \frac{\xi_{n}^2 v_{A}^2}{\nu_{in}+\nu_{en}}.
    \label{eq:Ambipolar}
\end{equation}
Ambipolar diffusion appears in the subsequent governing equations in the form of a Cowling diffusion coefficient, $\eta_C$, where $\eta_C = \eta_A + \eta$. 

The values of the collisional cross-sections necessary to calculate the above dissipative coefficients are taken to be $\sigma_{nn}=2.6\times 10^{-19}$ m$^{2}$, $\sigma_{in}=3.5 \times 10^{-19}$ m$^{2}$ and $\sigma_{en} = 10^{-19}$ m$^{2}$. In reality, these cross-sections are temperature dependent \citep[][]{VranjesKrstic2013} and, hence, should be spatially dependent, however, because of their weak variation with height and linear dependence on the dissipative coefficients, we treat them as constant quantities. Despite transverse inhomogeneity in the plasma density, we do not consider this to affect the discussed dissipative coefficients. Throughout our investigation we consider a quasi-neutral hydrogen plasma such that $n_e = n_i$ and here all number densities are given in m$^{-3}$ and temperature in K. We also consider elastic collisions between particles, so, momentum conservation requires that $\nu_{ab}m_an_a=\nu_{ba}m_bn_b$. 


With the inclusion of constant diffusion and viscosity coefficients, the linearized incompressible MHD equations reduce to a single governing equation describing the evolution of the magnetic field perturbation, $b$, as 



\begin{equation}
    \begin{split}
\frac{\partial^2 b}{\partial t^2} = & v_{A}^2(x)\frac{\partial^2 b}{\partial z^2} + \left[(\eta + \nu_{v})\frac{\partial^2}{\partial x^2} + (\eta_{C}+\nu_{v}) \frac{\partial^2}{\partial z^2}\right]\frac{\partial b}{\partial t} - \\ & -\nu_{v}\left[\eta\frac{\partial^2}{\partial x^2} + \eta_{C}\frac{\partial^2}{\partial z^2}\right]\nabla^2 b.   \label{eq:Gov_eq_full}
    \end{split}
\end{equation}

We employ a numerical program as detailed by \cite{McMurdo1} to solve the above governing equation in the presence of a pulse wave driver.


\section{Variation of Alfv\'en speed profiles}\label{sec:Alfven speed profiles}

One of the main aims of our investigation is to explore the variation in damping of an Alfv\'en pulse with a continuously excited wave. If a wave driver in the partially ionized region of the solar atmosphere operates for longer than a single period, the resulting wave attenuation achieved after propagating distances comparable to the chromospheric height is expected to closely resemble that of a continuously driven wave, due to the length scales involved. For this reason, we consider our driver to act for a single period only. We vary the frequency of the driver to achieve a varying initial wavelength along the magnetic field line of interest (this corresponds to the magnetic field line at the location of the maximal gradient in Alfv\'en speed profile), allowing for comparisons of the efficiency of damping with respect to the frequency of the driver.

\begin{figure}[htp]
   \centering
    \includegraphics[width=.45\textwidth]{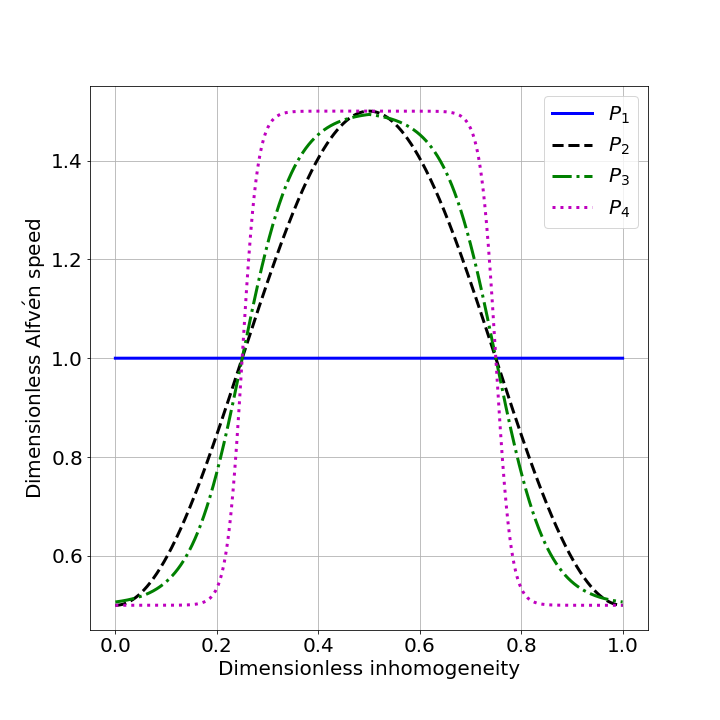}
    \caption{The different profiles of the Alfv\'en speeds as given by Equation (\ref{eq:vA_profiles}) used throughout the analysis are shown by curves of different colors. Speeds and lengths are given in dimensionless units. The $\tanh$ profiles are symmetric about the midpoint of the inhomogeneity in order to apply periodic boundary conditions in the transversal direction to our numerical solver, hence removing the effects of fixed boundaries.}
  \label{fig:vA_profiles}  
\end{figure}

To investigate the variability in damping of Alfv\'en waves caused by different levels of phase mixing in partially ionized plasmas, we prescribe that the simulated waves propagate at an Alfv\'en speed defined by one of the following four profiles, allowing for direct isolation of the effects of phase mixing, of varying steepness.

\begin{equation}
    \begin{split}
    & P_1: v_{A}(x) = v_{A1}, \\
    \\ & P_2: v_{A}(x) = v_{A1}\left(1 + \frac{1}{2}\cos\left[\frac{2\pi}{l_{inh}}(x - 0.5l_{inh})\right]\right),
    \\ & P_3: v_{A}(x) = v_{A1}\left(1 + \frac{1}{2}\tanh\left[\frac{x-0.25l_{inh}}{0.1 l_{inh}}\right]\right),
    \\ & P_4: v_{A}(x) = v_{A1}\left(1 + \frac{1}{2}\tanh\left[\frac{x-0.25l_{inh}}{0.03 l_{inh}}\right]\right),
    \end{split}
    \label{eq:vA_profiles}
\end{equation}
where $l_{inh}$ is the length scale of the inhomogeneity. In order to highlight the modifications in the damping of waves compared to a continuous driver employed by \cite{McMurdo1}, we once again choose $v_{A1}=20$ km s$^{-1}$, $l_{inh} = 300$ km, such that the Alfv\'en speed varies by a factor of $3$ across the inhomogeneity. This is achieved by assuming a density profile that varies by a factor of 9. Figure \ref{fig:vA_profiles} displays each of the four Alfv\'en speed profiles in dimensionless form. While profiles $P_2-P_4$ are dependent on the transversal coordinate, the profile $P_1$ represents a constant profile that will be used to highlight the effect of inhomogeneity in the Alfv\'en speed. 

The inhomogeneity length scales and characteristic Alfv\'en speeds necessary to perform our investigations were taken from previous studies involving propagating Alfv\'en waves in an inhomogeneous partially ionized plasma such as spicules and fibrils. Many observations have been able to quantify velocity amplitudes of Alfv\'enic waves \cite[e.g.,][]{OberservationHe,ObersvationOkamoto,ObservationGafeira,ObservationJafarzadeh,ObservationBate}. Quantifying the transverse density contrast from direct observations is very difficult, however, magnetoconvection codes have shown to produce such transversal density enhancements resembling fibrils and spicules and we chose the magnitude of the density contrast to be $9$, consistent with these simulations \citep[see e.g.,][]{SpiculeFormation,HalphaFormation,FibrilFormation,SpiculeFormation_AW}.

\subsection{Numerical limitations}\label{sec:NumericalLimitations}

\begin{figure}[htp]
    \centering
    \includegraphics[width=.45\textwidth]{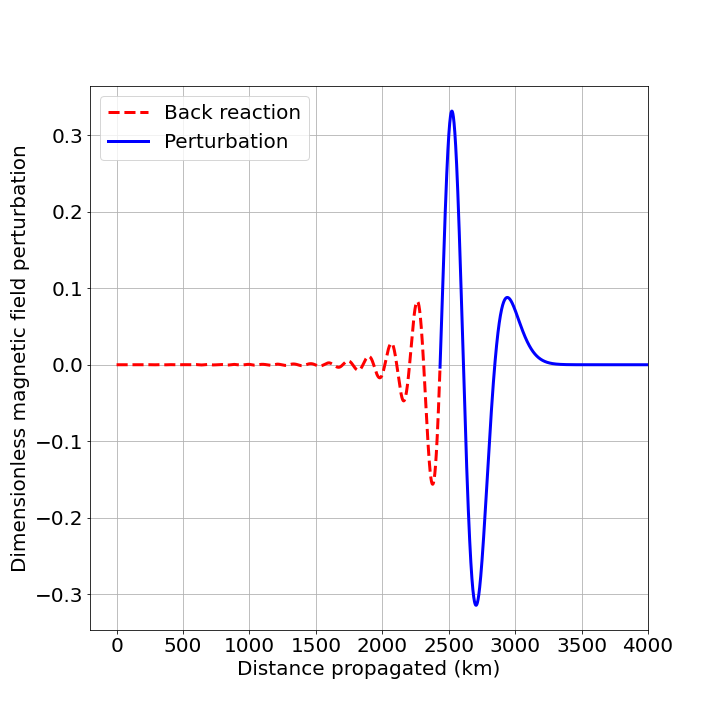}
    \caption{The profile of an Alfv\'en pulse excited using a pulse driver. The initial perturbation is colored in blue, while the high-frequency back-reaction is shown in red. The pulse was excited to a maximum dimensionless value of 1 and allowed to propagate until the end of the numerical domain. This simulation was performed for an ionization degree $\mu = 0.7645$, an Alfv\'en speed profile given by $P_3$, and a driver producing an initial wavelength of $300$ km.}
    \label{fig:5.4}
\end{figure}

For certain ionization degrees, we found that once the driver had been switched off, a high-frequency back-reaction propagates in the wake of the pulse that, based on simulations of varying resolutions, is a numerical artifact. Unlike numerical instability, this back-reaction does not grow with time or propagation. The various damping mechanisms we consider work to dissipate this numerical aspect of the solution. The amplitude of the back-reaction is limited by the presence of the various diffusive quantities present in the simulation. For simulations whose ionization degrees result in small dissipative coefficients, the amplitude of the back-reaction can be comparable to the amplitude of the initial perturbation, however, for simulations where the chosen ionization degree resulted in large dissipative coefficients, this back-reaction becomes negligible or does not exist at all for ionization degrees close to $\mu = 0.6$. While many numerical codes include an artificial diffusive quantity to deal with instabilities or anomalous numerical artifacts, we do not impose any such increased anomalous diffusion. Since this investigation is focused on how the damping of phase-mixed Alfv\'en pulses is affected by the ionization degree, we impose only the physical diffusive quantities predicted by formulae for that specific population of ions and neutrals given by the AL c7 model. To give the reader a visual representation of this effect, an example of this back-reaction is presented in Figure \ref{fig:5.4}, which was obtained by considering an ionization degree of $\mu = 0.7645$ (representing a neutrally dominated plasma), an Alfv\'en speed profile given by $P_3$, and a relatively high-frequency driver producing an initial wavelength of $300$ km (period of 15 seconds).
The appearance of the back-reaction can be attributed to the various truncation errors involved in the finite difference approximations and finite time step used in the fourth-order Runge-Kutta (RK4) time-stepping routine and is a direct consequence of turning off the driver. We note here that the changes in wave profile observed for the pulse driver that occur due to differing spatial and temporal resolutions are not seen in the case of the continuous driver. In order to avoid the influence of this back-reaction on the damping of waves, we consider ionization degrees where this back-reaction is negligible (meaning its amplitude is at most a few percent of the maximum amplitude of the main perturbation). 

\section{Results}\label{sec:Results}

For reasons explained in the previous section, we are going to consider a range of ionization degrees $\mu = 0.5036 - 0.6628$, which corresponds to the neutral population comprising approximately $2\% - 50\%$ of the total plasma population. These values ensure a sufficiently small numerical back-reaction, as discussed earlier. These values of ionization degree are, therefore, most relevant to the more highly ionized part of the solar chromosphere, rather than the photosphere.

In order to demonstrate the variation in damping profiles obtained from simulating an Alfv\'en pulse, we present the evolution of these waves at various time steps, over-plotting the normalized envelope of the continuously driven wave under the same conditions (i.e., identical Alfv\'en speed profile, ionization degree and frequency of wave driver), to align with the initial maximum and minimum of the profile of the pulse. At the start of the simulation, the initial profile of the magnetic field line is unperturbed, hence the first time step we plot corresponds to the moment the pulse has been perturbed for a single period. The profile of this pulse has already undergone some initial damping due to phase mixing and the various transport mechanisms begin working immediately upon perturbation of the magnetic field line. However, since the pulse perturbation is excited identically to the continuous driver, at least until the driver is turned off, the initial behavior of the finitely driven and continuously driven waves studied by \cite{McMurdo1} are identical. We begin by demonstrating the variation in behavior indicative of all our results through a range of examples, considering cases of strong phase mixing, no phase mixing, and simulations for which the ionization degree results in both strong diffusive quantities as well as weaker values.

\begin{figure*}
    \includegraphics[width=.33\textwidth]{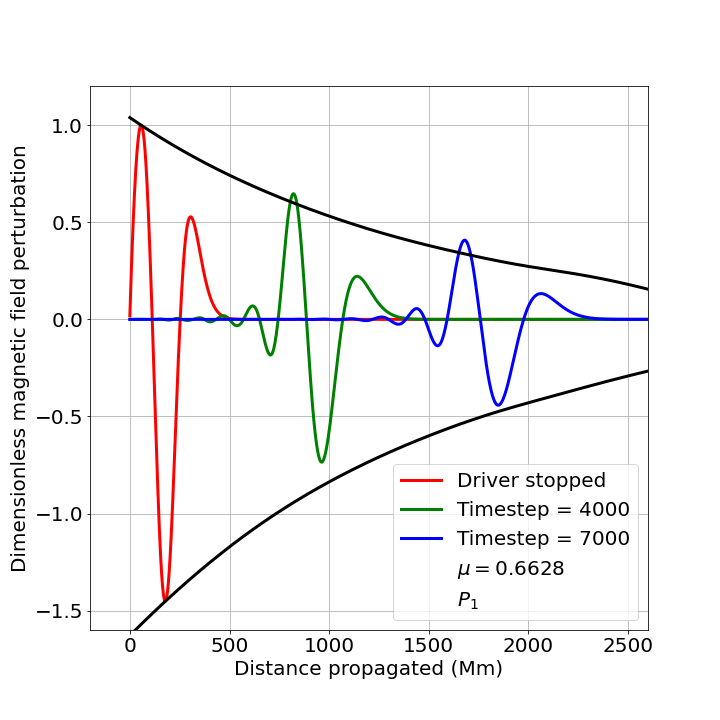}
    \includegraphics[width=.33\textwidth]{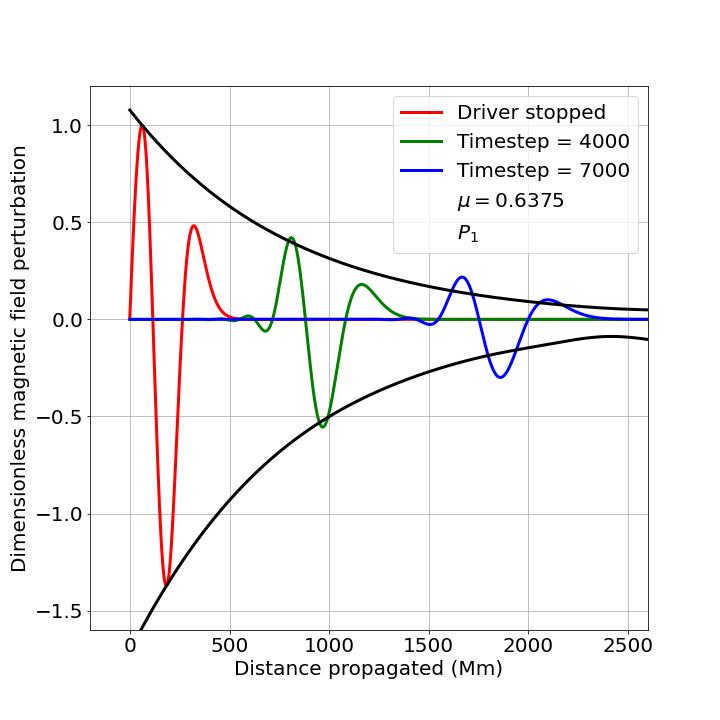}\includegraphics[width=.33\textwidth]{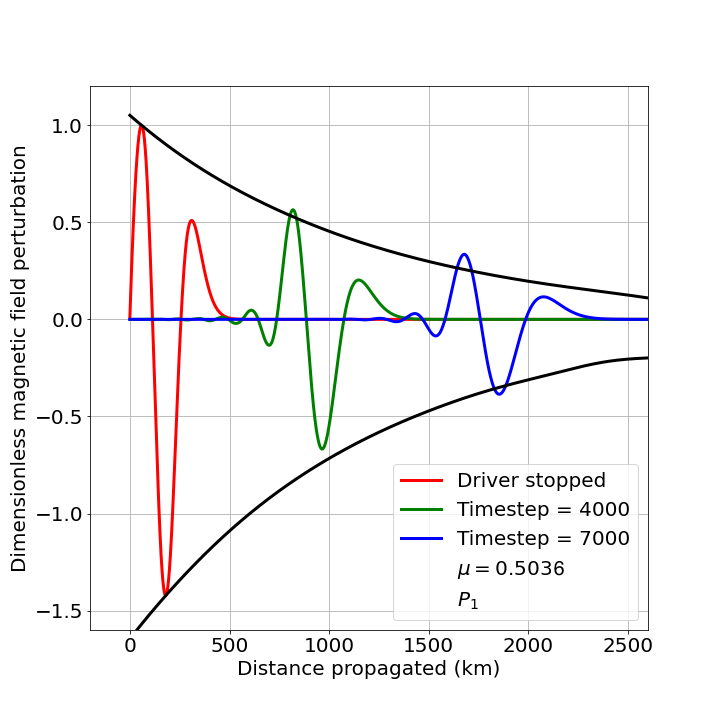}
    \\[\smallskipamount]
    \includegraphics[width=.33\textwidth]{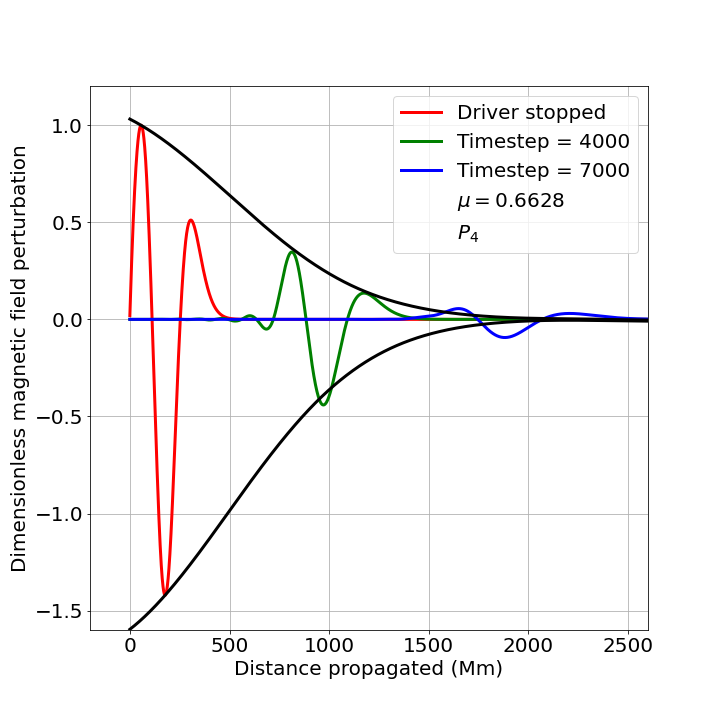}
    \includegraphics[width=.33\textwidth]{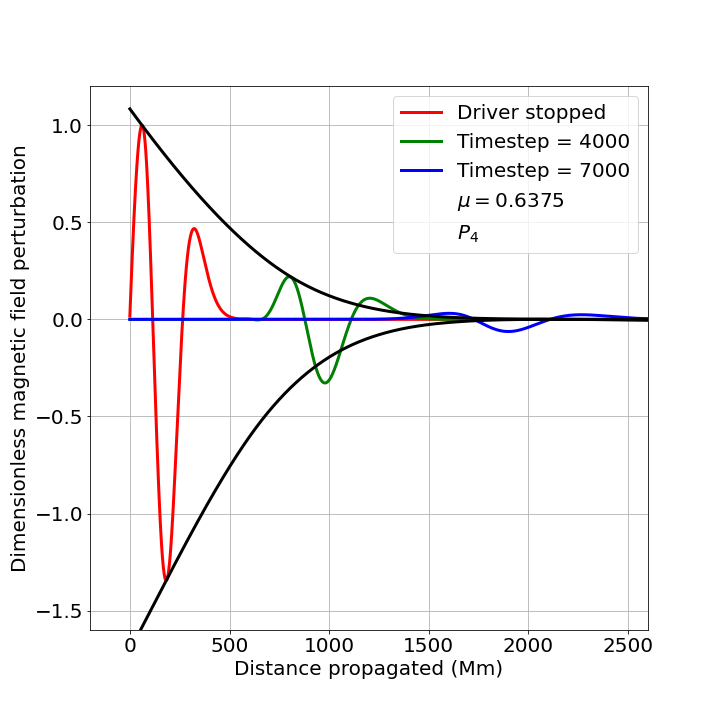}
    \includegraphics[width=.33\textwidth]{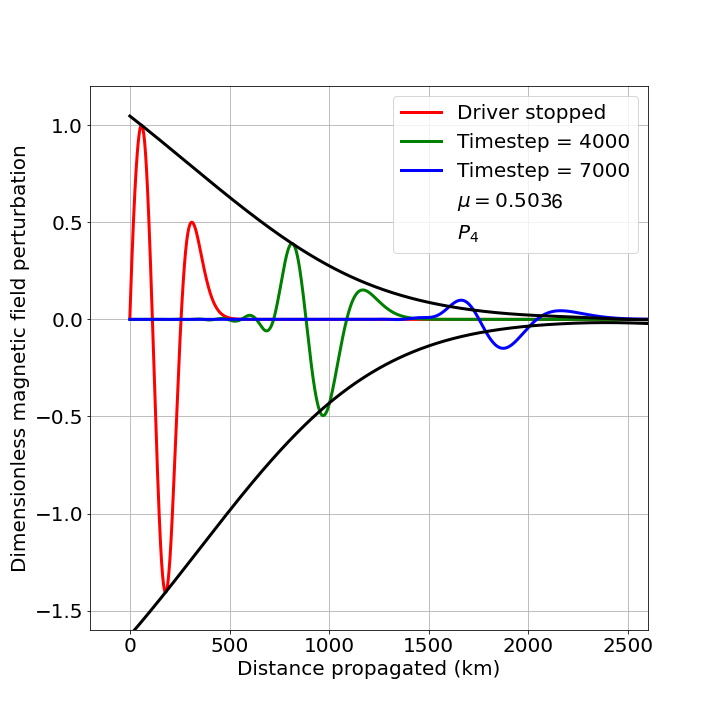}
    \caption{The evolution of a sinusoidally excited Alfv\'en pulse at three simulation time steps. The initial profile corresponds to the moment the driver terminates and represents a pulse with a wavelength of $300$ km (shown in red at the base of the domain). Subsequent time steps (shown in the middle of the domain in green and at the end of the domain in blue) reveal the evolution in the wave profile. Additionally, the envelope of the continuously excited Alfv\'en wave, generated under identical conditions, is superimposed for comparison (black line). Differences arise solely due to the finite lifetime nature of the pulse, which lacks a continual energy injection at the domain's base. Here the two rows correspond to wave propagating with Alfv\'en speed profiles given by the $P_1$ (top row) and $P_4$ (bottom row) profiles.}
    \label{fig:Continuous and finite}
\end{figure*}

Figure \ref{fig:Continuous and finite} shows that the continuously driven wave exhibits more efficient damping for all ionization degrees considered since the profile of the pulse exceeds the amplitude of the envelope of the continuous driver. As the pulse profile expands, its effective wavelength increases such that the damping due to longitudinal gradients (predominantly Cowling diffusion) becomes less effective. In the case of the continuous driver, the persistent energy injection at the base of the domain works to preserve the wavelength of the Alfv\'en wave subject to phase mixing. This effect is not seen in the case of the finite-lifetime driver, as there are no preceding waves. 

The ionization degree influences various physical aspects of the results. It determines the values of the dissipative coefficients, which in turn affect the system's behavior. Higher values of the dissipative coefficients associated with cross-field derivatives ($\eta$ and $\nu_v$) lead to increased pulse broadening due to the enhanced effects of phase mixing. Similarly, increased dissipative coefficients related to longitudinal gradients ($\eta_C$ and $\nu_v$) result in wavelength-dependent damping. For pulses, the effective wavelength increases due to pulse broadening driven by phase mixing, that is causing weaker damping compared to the continuous case since now the longitudinal gradients are reducing, meaning the waves damp as if they have larger wavelengths than their initial state, resulting in weaker damping. This implies that for waves propagating in the presence of an ionization degree which results in large dissipative coefficients, the behavior of a pulse and that of a continually driven wave differ the most i.e., for ionization degrees approximately (or close to) $\mu = 0.6$. 

The differences we refer to here are not purely evident in the amplitude of the pulse. Due to the broadening of the pulse, the subsequent reduction in amplitude and larger effective wavelength are competing effects. When comparing only the amplitude of the pulse, this can often be seen to balance one another out, leading one to potentially falsely conclude that pulses and continuously driven waves damp similarly. In the cases of ionization degrees closest to $\mu = 0.6$, the dissipative terms in Equation (\ref{eq:Gov_eq_full}) are at their largest because the dissipative coefficients related to such derivatives become very large. This means that the change in effective wavelength of the pulses are more exaggerated compared to the case of an ionization degree that results in weaker dissipative coefficients. More specifically, the closer the ionization degree is to $\mu = 0.6$, the more the pulse will broaden. This effect can be seen when comparing the differences in width of the pulses in the top row compared with the bottom row of Figure \ref{fig:Continuous and finite}. The pulses that experience phase mixing have broadened by 50\%, 78\% and 32\% as we go from the left panel through to the right panel, respectively, when comparing with the width of the pulses not subject to phase mixing. If the plasmas ionization degree was close to either $0.5$ or $1$, then the discrepancies in the pulse width would be much less due to the smallness of the dissipative coefficients.



The slight widening effect we see in the case of the homogeneous Alfv\'en speed (top row of Figure \ref{fig:Continuous and finite}) is due to the truncation errors involved in the finite difference approximations used throughout the numerical solver, whereas in the case of waves propagating in the presence of our steepest Alfv\'en speed profile (second row of Figure \ref{fig:Continuous and finite}), the additional widening over the homogeneous case occurs due to phase mixing. In order to ascertain the importance of the Alfv\'en speed profile on the attenuation of these pulses, we now present further results of a comparative study between simulations adopting a homogeneous Alfv\'en speed profile and those with inhomogeneous profiles. 


\subsection{Variation in damping due to Alfv\'en speed profile}

\begin{figure*}
    \includegraphics[width=.33\textwidth]{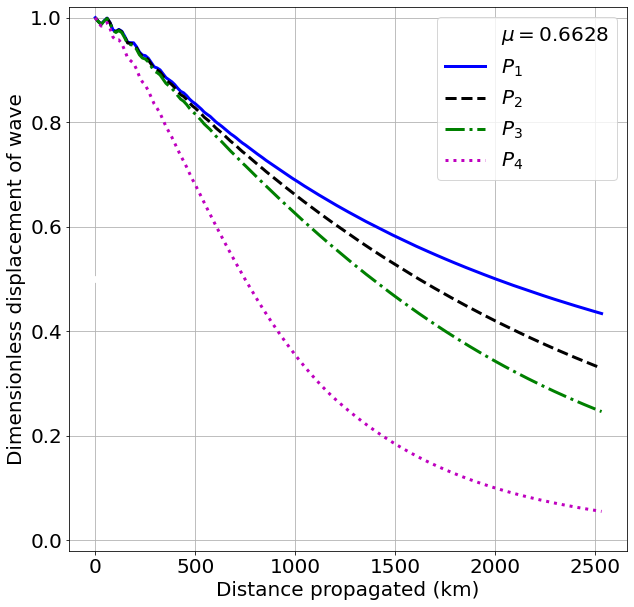}
    \includegraphics[width=.33\textwidth]{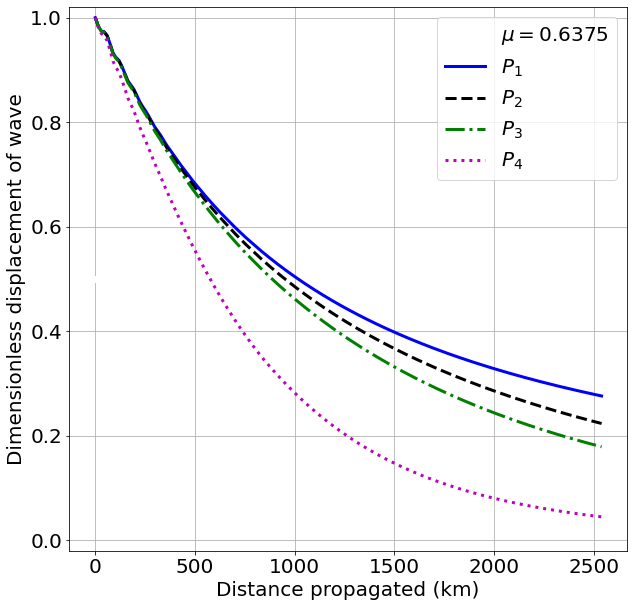}    \includegraphics[width=.33\textwidth]{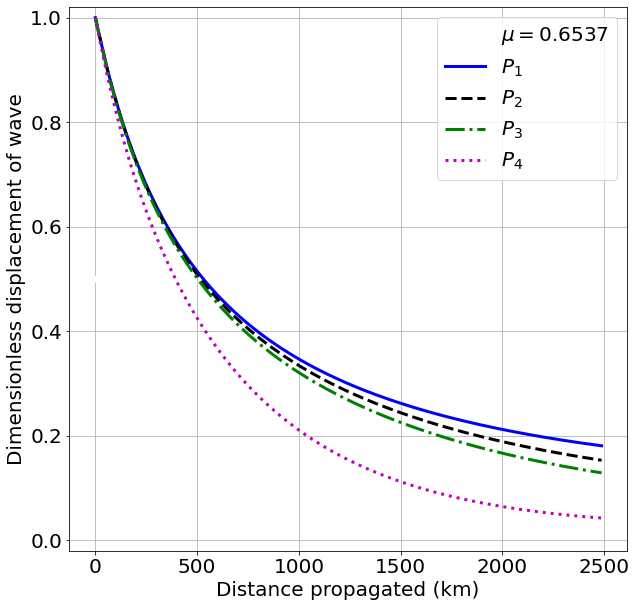}\hfill
    \\[\smallskipamount]
    \includegraphics[width=.33\textwidth]{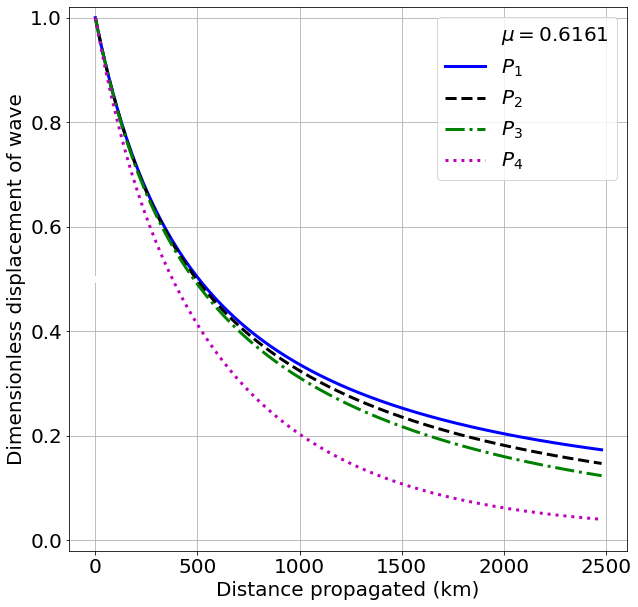}
    \includegraphics[width=.33\textwidth]{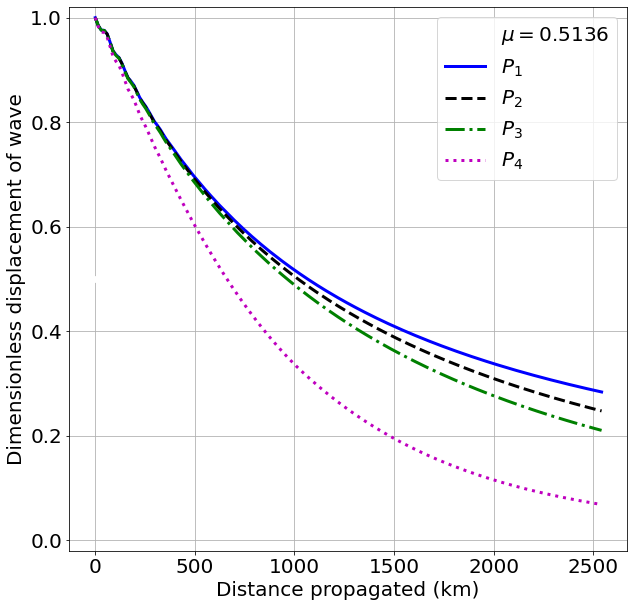}\hfill
    \includegraphics[width=.33\textwidth]{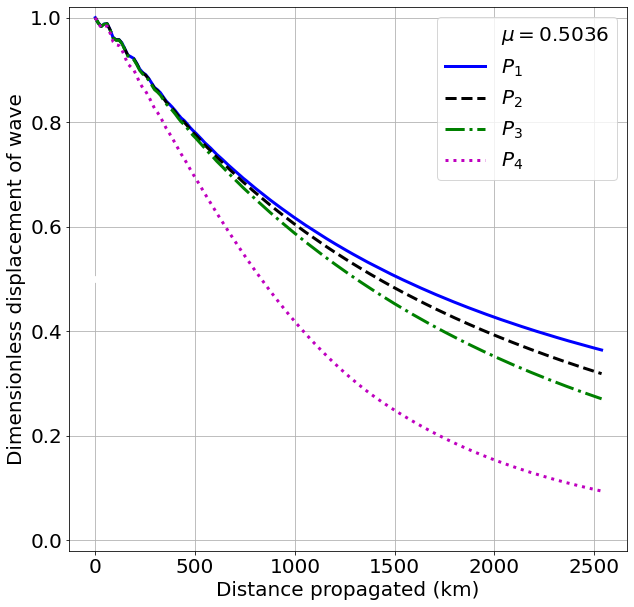}\hfill
    \caption{The variation in the displacement of each of the Alfv\'en pulses with propagation for the four different Alfv\'en speed profiles given by $P_1 - P_4$ profiles (shown by different colors), for six different ionization degrees and an initial wavelength of $300$ km.}
    \label{fig:Area change for vAs}
\end{figure*}

Phase mixing works to enhance the dissipative mechanisms associated with cross-field derivatives. Due to the finite nature of the considered driver, the interactions between adjacent waves propagating at differing speeds now cause an increase in the width of the pulse, since there is no constant injection of energy at the base of the domain working to regulate the wavelength throughout the simulation. Due to this widening of the pulse with propagation, it is insufficient to use the amplitude of the wave as a proxy for the energy remaining within the disturbance, instead, we measure the variation in the total displacement of the wave to provide a more accurate representation of the dissipation of energy with propagation. The displacement of a wave is defined as the area of the absolute value of the wave's profile bound by the $z$-axis along which it propagates. Figure \ref{fig:Area change for vAs} shows the variation in displacement for ionization degrees in the range $\mu = 0.5036 - 0.6628$ with an initial effective wavelength of $300$ km along the magnetic field line that aligns with the location of the maximal gradient in Alfv\'en speed, where the Alfv\'en speed profiles are distinguished by different colors. 

These results show that waves propagating in the presence of a steep Alfv\'en speed profile reduce their total displacement far more effectively than those in a more weakly inhomogeneous ($P_2 - P_3$) and all more than the homogeneous profile ($P_1$). While remnants of the perturbation are expected to protrude into the higher layers of the solar atmosphere, waves propagating in the presence of a strongly inhomogeneous Alfv\'en speed still deposit the majority of their wave energy before propagating a distance comparable to the height of the chromosphere ($2,500$ km). Clearly, throughout all simulations displayed in Figure \ref{fig:Area change for vAs}, there is a distinct difference between the homogeneous case ($P_1$) and the most weakly varying inhomogeneous case ($P_2$). The effects seen in Figure \ref{fig:Continuous and finite} of the pulse widening in the case of the homogeneous profile, are understood to be small due to this distinct difference between the blue (solid) lines ($P_1$) and the black (dashed) lines ($P_2$) in Figure \ref{fig:Area change for vAs}. The reduction in damping that occurs between the top-left, top-middle and top-right panels in Figure \ref{fig:Area change for vAs}, shows an initial decrease in ionization degree, followed by an increase. This effect occurs due to the multi valued nature of the ionization degree as plotted with height in Figure \ref{fig:IonisationDegree_Temp}. 


\subsection{Damping profiles}\label{sec:DampingProfiles}

By performing a Fourier analysis, it is clear that in a homogeneous plasma, continuously driven waves damp exponentially due to various resistive quantities. In the presence of inhomogeneity, \cite{HeyvaertsPriest1983} found that continuously driven phase-mixed Alfv\'en waves decay exponentially, but their damping length now depends on both the non-ideal effects and the transverse inhomogeneity. Including partial ionization effects \cite{McMurdo1} found that (in leading order), the damping profile of continuously driven waves propagating in an inhomogeneous environment could be represented by the sum of two exponential functions, one recovering the work of \cite{HeyvaertsPriest1983} and the other occurring due to the Cowling diffusion, i.e., the presence of neutrals. \cite{Hood2002} found that a pulse propagating in the presence of inhomogeneity decayed algebraically, rather than exponentially. This result occurs, not due to phase mixing, but rather because a pulse can be represented as the sum of many continuously driven waves, each damping exponentially. The result of this is that the decay of a pulse obeys a power law that depends on the various simulation parameters. 

\begin{figure}
    \centering
    \includegraphics[width=0.45\textwidth]{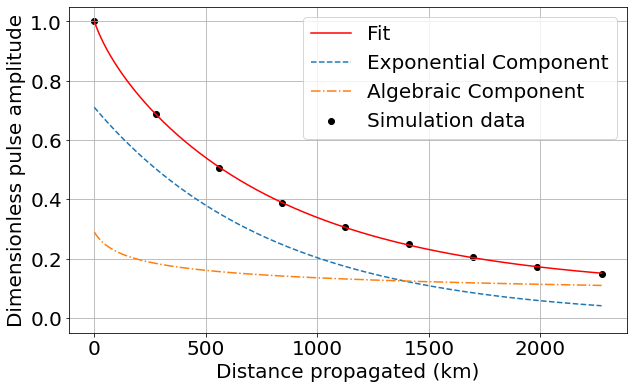}
    \includegraphics[width=0.45\textwidth]{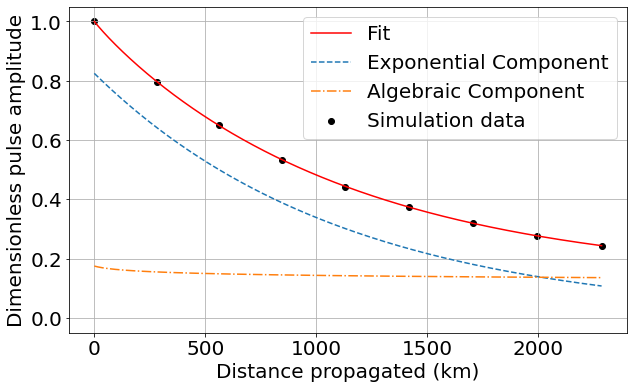}
    \caption{The damping profile for an Alfv\'en pulse is shown to be a sum of an exponential and algebraic component. The relative importance of each damping profile varies with propagation as evidenced by the intersection of these profiles. The physical parameters for this simulation consist of an ionization degree $\mu = 0.5036$ (bottom panel) $\mu = 0.6375$ (top panel), an Alfv\'en speed profile given by $P_1$ and an initial wave length of $300$ km.}
    \label{fig:Change_damping_profile}
\end{figure}

\begin{figure*}
    \includegraphics[width=.225\textwidth]{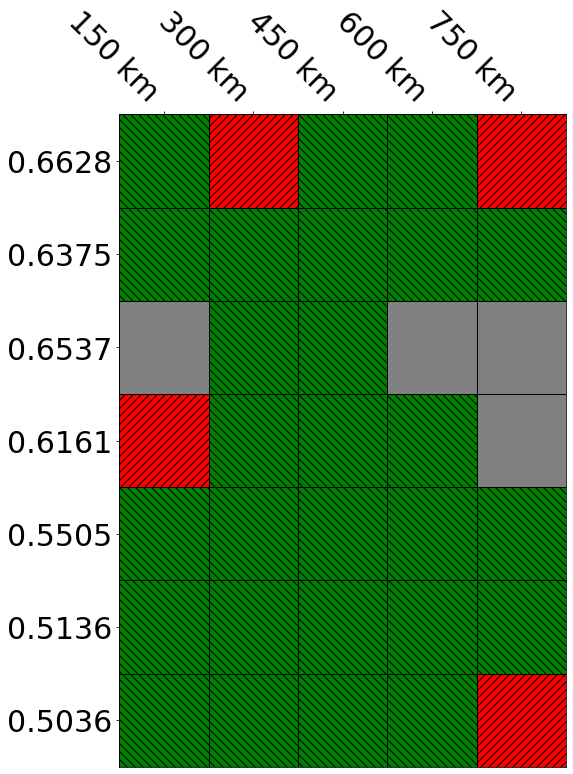}\hfill
    \includegraphics[width=.225\textwidth]{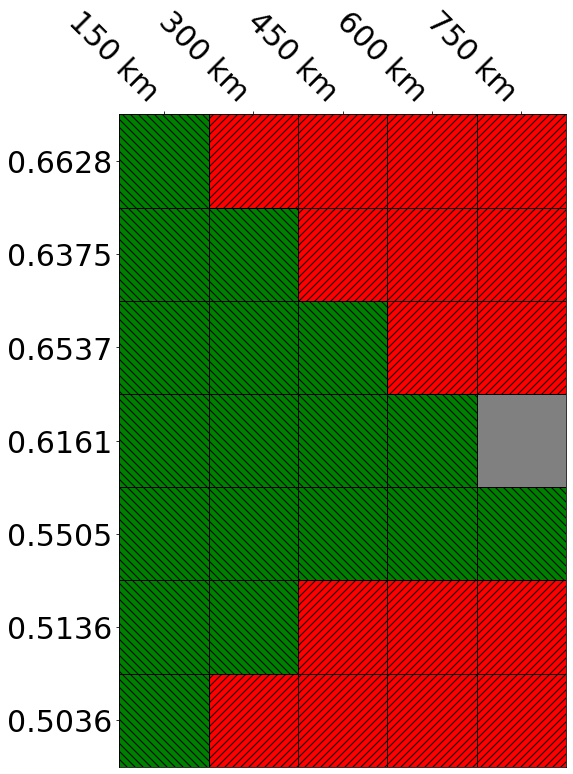}\hfill
    \includegraphics[width=.225\textwidth]{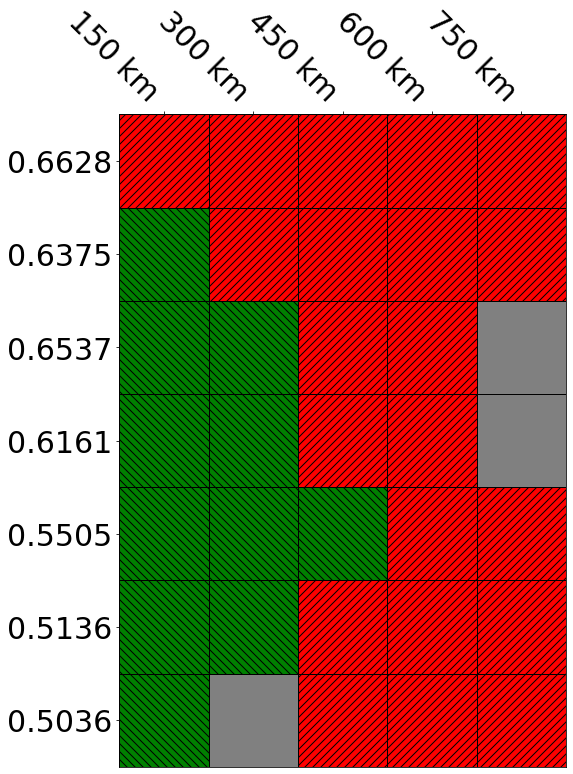}\hfill
    \includegraphics[width=.225\textwidth]{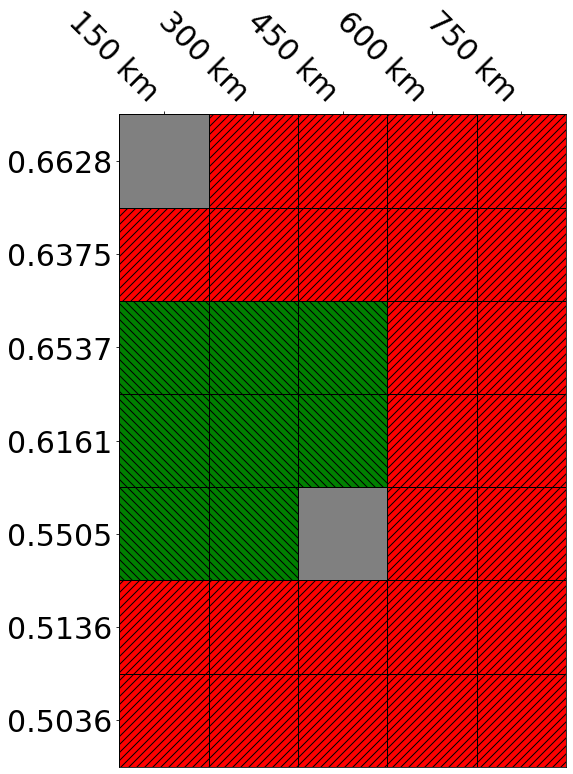}\hfill
    \caption{For each simulation, the resulting damping profiles are represented using a color-coded grid: green indicates a change in the damping profile, red signifies no change, and gray represents cases where the results were inconclusive or no clear change could be determined. The $x$-axis of the grid represents the variation in the initial wavelength of the pulse, while the $y$-axis of the grid represents the variation in ionization degree considered throughout the simulation. The variation between figures from left ($P_1$) to right ($P_4$) occur due to a variation in the Alfv\'en speed profile.}
    \label{fig:Damping_change_diagram}
\end{figure*}

When studying the damping profiles of the simulated kink waves, \cite{Pascoe2012} observed a change in damping profiles and it was shown that the damping of kink waves was initially well approximated by a Gaussian function, before transitioning to an exponential profile after some empirically determined distance. In our study we observed a similar change in the damping profile for the simulated pulses. It was evident that a change in the damping profile was dependent on various physical parameters used throughout the simulation, specifically the ionization degree, Alfv\'en speed profile and driver frequency. By fitting a combination of functions, it was found that the decay of these pulses was initially well approximated by an exponential function, but for later times (larger distances propagated), the profile shifted towards an algebraic decay. An example is shown in Figure \ref{fig:Change_damping_profile}. We emphasize that the onset of a secondary damping regime is not a direct consequence of phase mixing. Indeed, as shown in Figure \ref{fig:Damping_change_diagram}, phase mixing can inhibit changes in the damping profile as transverse inhomogeneity and hence phase mixing maintains the coherence of the pulse’s original driving frequency. Instead, we ascribe this effect to partial ionization, which introduces additional damping mechanisms leading to an additional damping regime once the pulse has propagated beyond a certain distance.

For an infinite wave train propagating in a homogeneous plasma driven at a single frequency, a straightforward Fourier analysis predicts an amplitude decay proportional to $e^{-2 \eta_C k^2 t}$, where $k$ is the wavenumber. A pulse can be thought of as a superposition of many infinite wave trains of different frequencies, where Fourier analysis used to obtain the above damping rate is valid for each of the constituent components of the pulse (with varying wavenumber, $k$). In this case, if all constituent frequencies had comparable contributions to the overall behavior of the pulse, one would expect the decay rate to behave according to a power law (by summing together the various exponential decaying frequencies). In other words, the decay would be algebraic. 

Each pulse in our simulations was driven with a single frequency, resulting in the initial damping behavior being well approximated by an exponential function. However, with propagation, the interactions with neighboring waves results in the contribution of waves with differing wavelengths. Therefore, at later times (or equally, after the pulse has propagated further), the damping of the pulse becomes better approximated by an algebraic function. The effect of a change in damping profiles is seen in fewer simulations as the inhomogeneity in Alfv\'en speed becomes more severe since phase mixing kills off the higher frequencies faster relative to the lower frequencies. Phase mixing, therefore, preserved the relative magnitude difference between the dominant frequency and the minor frequency contributions resulting in the decay of the signal being well approximated by exponential decay. This is because phase mixing works more effectively to damp high-frequency perturbations compared to lower frequency perturbations. For simulations where a change in damping profiles was observed there was, at later times, a more balanced contribution of wave components each decaying exponentially leading to a resultant algebraic decay.


For Alfv\'en pulses propagating at ionization degrees closest to $\mu = 0.6$ in environments that were least inhomogeneous, the wave damping profiles (obtained by tracking the maximum value of the perturbation as a function of time) initially behaved as a decaying exponential, then, beyond some given distance, it changed to algebraic decay. This result is analogous to that found by \cite{Pascoe2012,Pascoe2013,Pascoe2015}. In this investigation, we found that the presence of steeper inhomogeneities such as those given by $P_3$ and $P_4$ \citep[steeper than those employed by][]{Hood2002} ensures an exponential decay rate for longer distances, given the same driver frequency and ionization degree, overriding the change in algebraic decay which we observe in the homogeneous simulations, often resulting in no change in decay profiles.

For each simulation, a qualitative assessment was performed to determine the presence of any changes in the damping profiles and in Figure \ref{fig:Damping_change_diagram} we present four diagrams representing whether a change was observed. We considered a variety of wave frequencies and ionization degrees for each of the four Alfv\'en speed profiles.

\subsection{Variation in damping due to frequency}

In the present study, we investigated five different frequencies of wave driver, with dimensional values varying between $27-133$ mHz, which, prior to any stretching of the profiles of the waves (due to phase mixing), produce wavelengths between $150 - 750$ km. We choose again to present the variation in the total displacement of the wave rather than the amplitude to distinguish between the effects of phase mixing on widening the pulse and the reduction in the perturbation (representative of the energy left in the wave).  

\begin{figure*}
    \includegraphics[width=.45\textwidth]{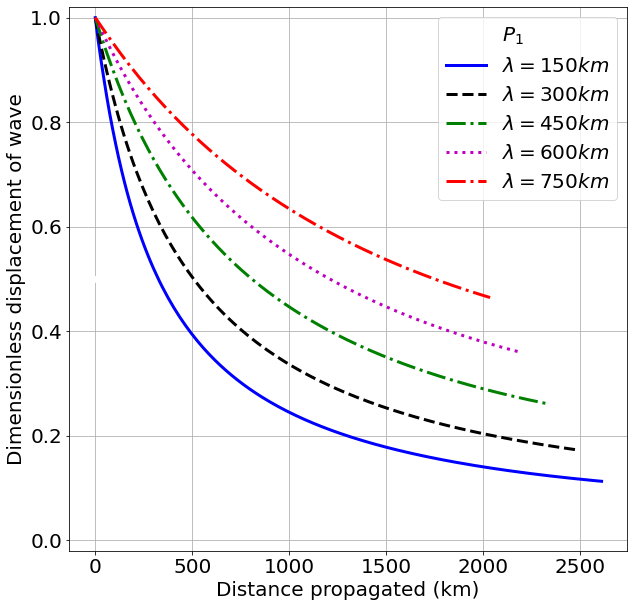}\hfill
    \includegraphics[width=.45\textwidth]{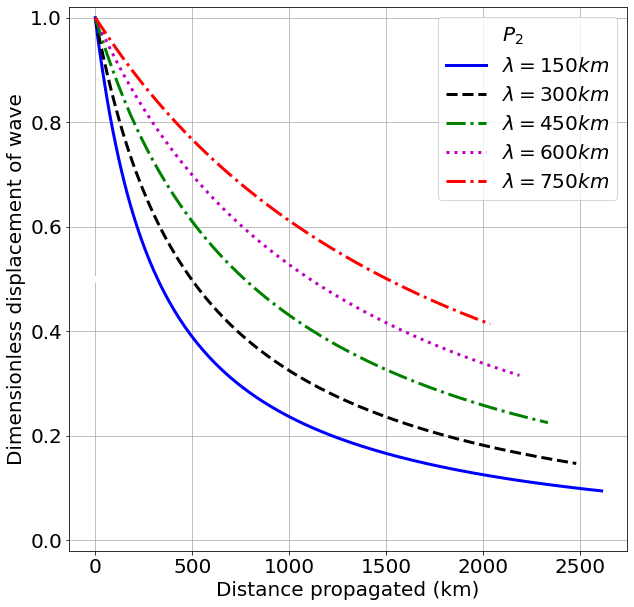}\hfill
    \\[\smallskipamount]
    \includegraphics[width=.45\textwidth]{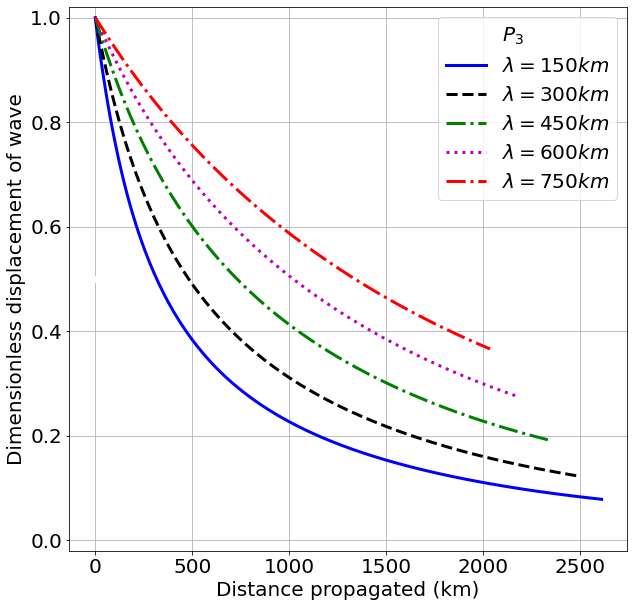}\hfill
    \includegraphics[width=.45\textwidth]{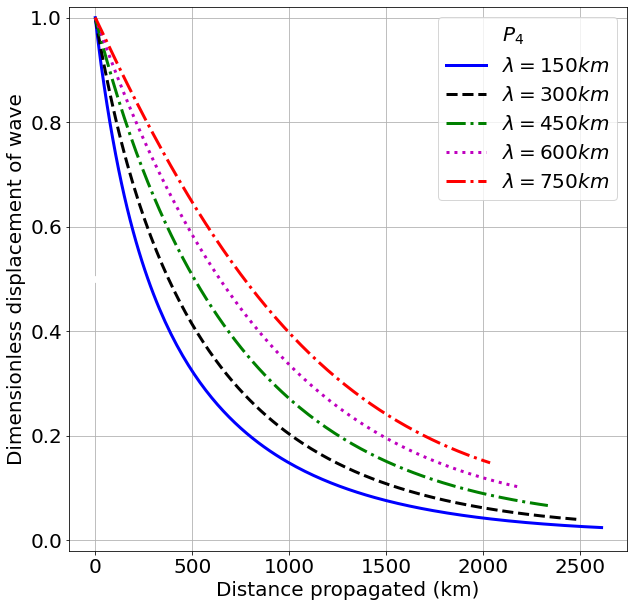}\hfill
    \caption{The variation in the displacement of finitely excited Alfv\'en waves with distance for each of the four Alfv\'en speed profiles considering a single ionization degree given by $\mu = 0.6161$.}
    \label{fig:Area change for wavelength}
\end{figure*}

The results of our simulations are shown in Figure \ref{fig:Area change for wavelength}, where we calculate the distance propagated by the wave as the distance the center peak of the perturbation has propagated from the moment the driver is turned off. Simulations are terminated once the perturbation reaches the boundary of our numerical domain, in order to avoid effects corresponding to wave reflection. The center peak was chosen to avoid the influence of an elongated wave profile in the presence of steep Alfv\'en speed gradients. The reason for the increased distance with reduced initial wavelength can be easily explained. The middle peak occurs earlier in the domain for the shorter wavelengths allowing for the simulation to persist longer. Furthermore, the displacement of pulses with larger effective wavelengths is naturally larger than those with shorter wavelengths, hence, we normalize the total displacement of each pulse such that the initial displacement is one for all simulations. Since we work in a linear regime, we will not see any variation in wave propagation due to the wave amplitude. Figure \ref{fig:Area change for wavelength} clearly demonstrates that pulses with a shorter initial wavelength (higher frequency) reduce their total displacement far more effectively than those with larger wavelengths; this result is analogous to the conclusions drawn by \cite{McMurdo1}, where shorter wavelength waves (or similarly, higher frequency waves) damp more effectively than their longer wavelength counterparts. Pulses with the shortest wavelengths dissipate their wave initially very quickly, while pulses with larger wavelengths decay at a much steadier rate. In the presence of the homogeneous Alfv\'en speed, it is only the shortest of wavelengths that dissipate their energy over the distance comparable to the height of the chromosphere, while for the steepest Alfv\'en speed profile, all wavelengths dissipate over 80 \% of their wave energy within 2000 km, highlighting the efficiency of phase mixing in damping Alfv\'en waves in partially ionized solar plasmas. 

\begin{figure*}
    \includegraphics[width=.33\textwidth]{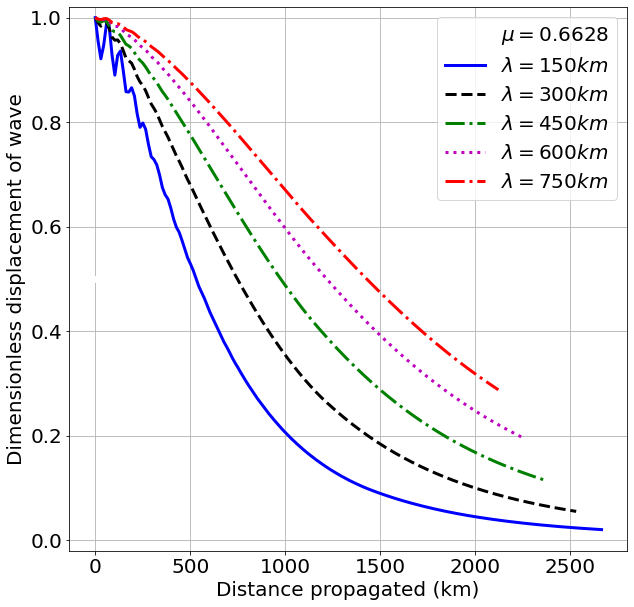}
    \includegraphics[width=.33\textwidth]{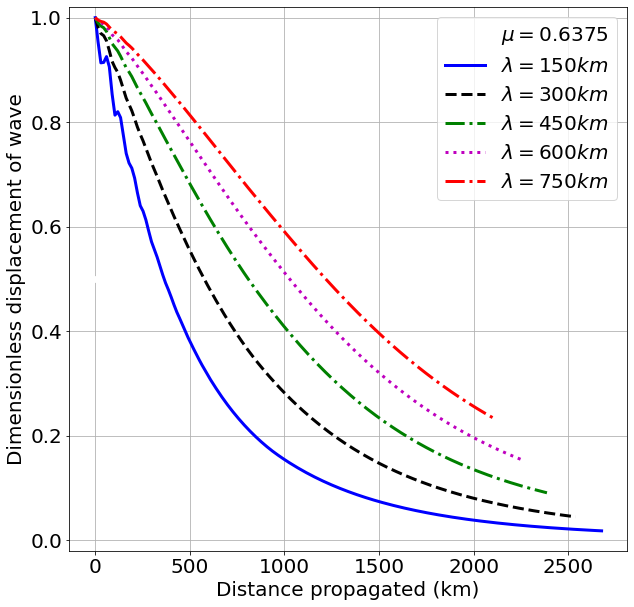}
    \includegraphics[width=.33\textwidth]{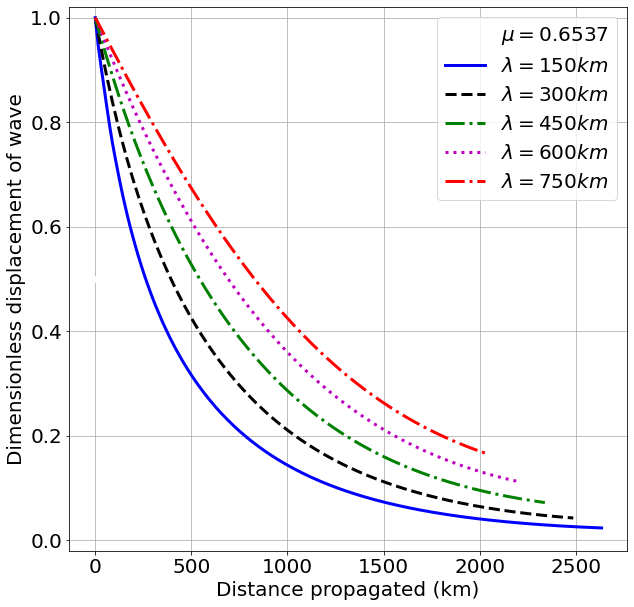}\hfill
    \\[\smallskipamount]
    \includegraphics[width=.33\textwidth]{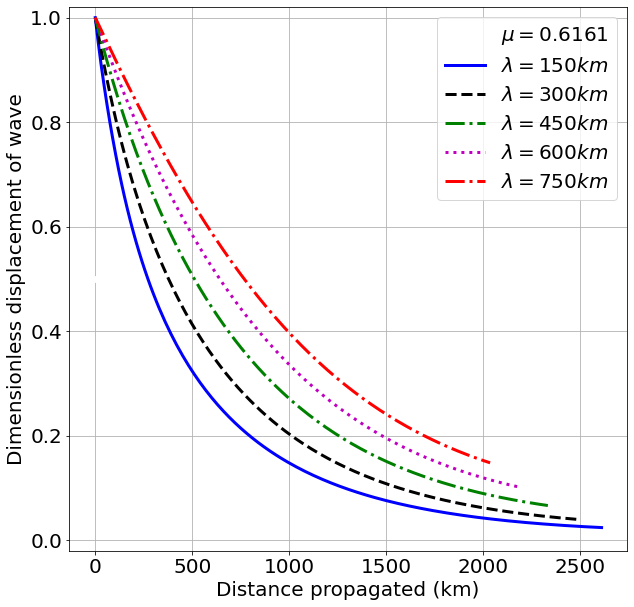}
    \includegraphics[width=.33\textwidth]{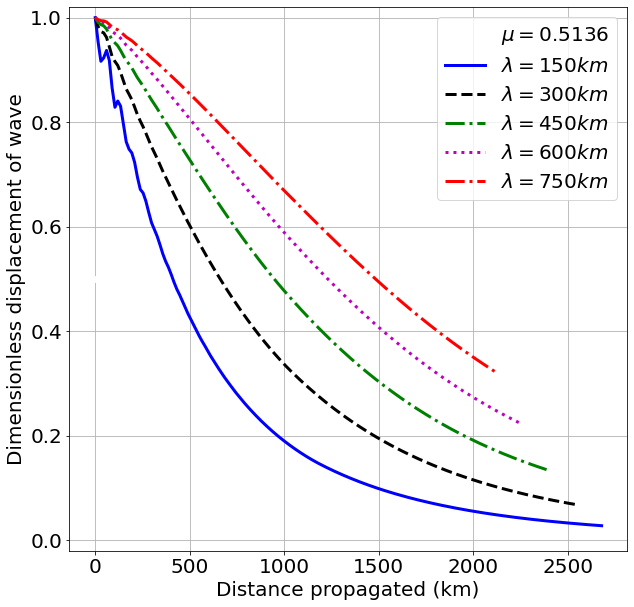}
    \includegraphics[width=.33\textwidth]{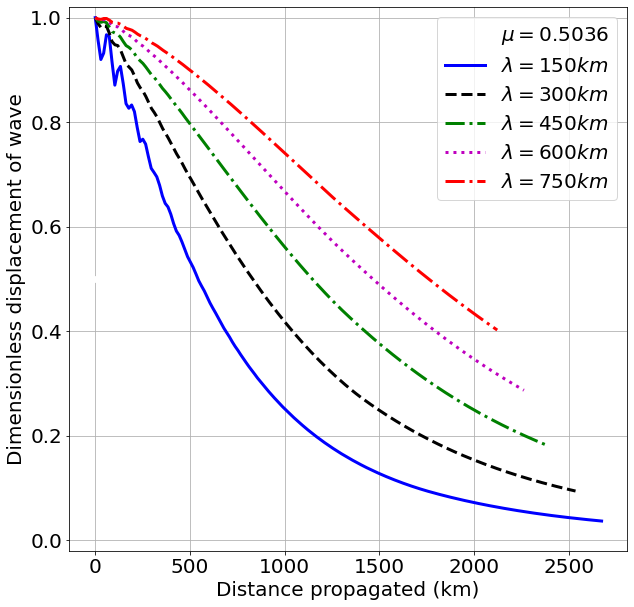}
    \caption{The variation in the displacement of our Alfv\'en pulses with distance for the five different wavelengths for six different ionization degrees in the case of the Alfv\'en speed profile given by $P_4$.}
    \label{fig:5.8}
\end{figure*}

The variation in the dimensionless displacement of the wave with distance for various ionization degrees is shown in Figure \ref{fig:5.8}. The oscillatory behavior of the displacement evident in ionization degrees at the extreme values considered here is a numerical phenomenon, arising from deactivating the wave driver within a weakly diffusive plasma. This effect is most noticeable in the case of a high-frequency driver, due to the increased truncation errors relative to the wavelength. Examining the results, we do not expect this numerical effect to play a major role in the qualitative conclusions we draw. An extreme case of this has already been seen in \ref{sec:NumericalLimitations}. We present all results in terms of the dimensionless transverse displacement because it is directly observable. To estimate the wave energy at a given time, $t_1$, we assume equipartition between kinetic and magnetic energy contributions and define the normalized pulse energy along a field line at $x=x_0$ by 
\begin{equation}
    E(t_1,x_0) = \frac{\int_{0}^{z_{\mathrm{max}}}b(x_0,z,t_1)^2\mathrm{d}z}{\int_{0}^{z_{\mathrm{max}}}b(x_0,z,t_0)^2\mathrm{d}z},
    \label{eq:wave_en}
\end{equation}
where $z_{\mathrm{max}}$ is the maximum height of the domain and the denominator calculates the energy in the wave the moment the driver is turned off such that the energy is normalized to one (in dimensionless units). Our aim is to examine the evolution of $E$ over time. Constant quantities are not shown in Equation (\ref{eq:wave_en}) as they cancel during the division. Although not exact in a mathematical sense, we find that the square of the pulse’s displacement reproduces the decay of \(E\) to high accuracy. Consequently, the energy damping rate scales approximately as the square of the perturbed area, implying that the pulse’s energy decays significantly more rapidly than its displacement amplitude.

With respect to the effect the ionization degree has on the attenuation of Alfv\'en pulses, it is easy to see that Alfv\'en pulses propagating in a plasma with an ionization degree close to $\mu = 0.6$, result is the most effective damping, while deviation towards either a more ionized or less ionizes plasma results in lesser attenuation. 

\begin{figure}
    \centering
    \includegraphics[width=0.45\textwidth]{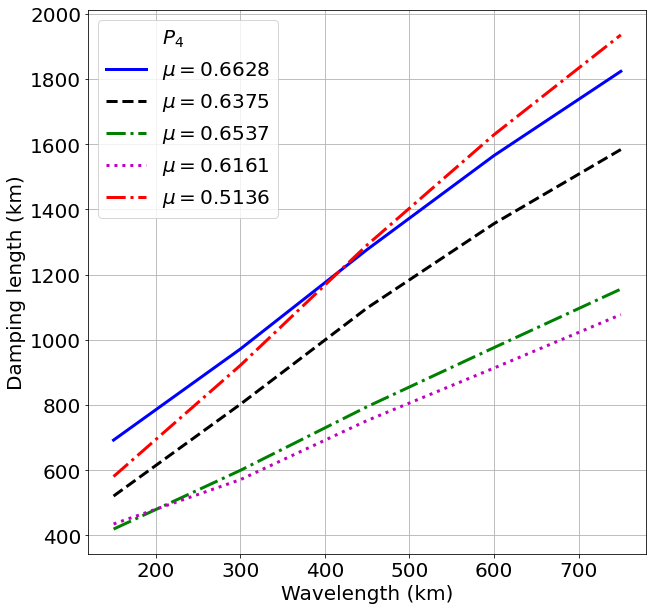}
    \includegraphics[width=0.45\textwidth]{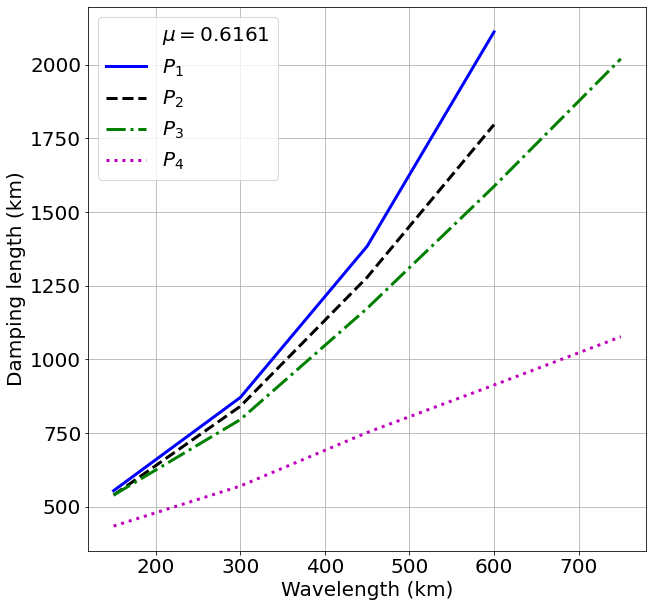}
    \caption{Damping length of the integrated displacement of the wave shown as functions of wavelength. For $P_{1,2}$ the damping length is not achieved for the longest wavelength and has to be omitted.}
    \label{fig:Damping_length}
\end{figure}

We define the damping length as the distance over which the displacement area of the Alfv\'en pulse decreases by a factor of $e$. This definition is used to quantify the widening of the pulse due to phase mixing rather than considering the change in the amplitude as representative for the damping process. Figure \ref{fig:Damping_length} shows the dependence of this damping length on wavelength for various ionization degrees (upper panel), and  wavelengths (through a varied frequency driver) and Alfv\'en speed profiles (lower panel).

As the top panel of Figure \ref{fig:Damping_length} demonstrates, there is a linear relationship between the damping length of the pulse and the wavelength of the wave for all ionization degrees studied in the strong phase mixing case. Albeit we measure the displacement of the pulse rather than the amplitude, this result is somewhat analogous to the result obtained earlier by many studies. Indeed, \cite{Terradas2010} demonstrated that for low frequency kink waves propagating in coronal loops, the damping length due to resonant absorption is inversely proportional to the frequency of waves, implying a direct proportionality between damping length and wavelength. This relationship (known as the TGV relation) suggests that high-frequency (short wavelength) waves are more efficiently damped causing solar magnetic waveguides to act as natural low-pass filters. Furthermore, a recent study by \cite{Tiwari2021} found a linear dependence between the damping length of kink waves and the period waves. Moreover, the study by \cite{Rawat2023} found that the damping lengths of slow magnetoacoutic waves in sunspot fans increases with the wave period, indicating a frequency-dependent damping behaviour. The various gradients of the lines in the top panel of Figure \ref{fig:Damping_length} occur due to the variation in dissipative coefficients. Some ionization degrees result in dissipative coefficients related to longitudinal gradients preferentially increasing while in others the transverse gradients lead to an ability to damp larger or shorter wavelengths more effectively. This linear dependence is not observed for all levels of phase mixing as can be seen in the bottom panel of Figure \ref{fig:Damping_length}, whereby weaker phase mixing leads to an apparent exponential increase in damping length. This suggests that strong gradients in the Alfv\'en speed still ensure that small enough scales are achieved in large wavelength waves to result in strong damping. We note that in our simulations we consider a constant Alfv\'en speed throughout each simulation and hence it is simple to convert between the wavelength and the period, by dividing the horizontal axis by $v_{A1}$.



\subsection{Heating rates}

To investigate the efficiency of Alfv\'en pulses at heating the surrounding plasma through phase mixing, we conduct a comparative study of the heating rates obtained for the continuously driven and the finitely driven Alfv\'en waves. The resistive heating rate for phase-mixed Alfv\'en waves is given by 

\begin{equation}
    \begin{split}
        Q = Q_{res} = & \frac{1}{\mu_{0}}\left[\eta\left(\frac{\partial b}{\partial x}\right)^2 + \eta_{C}\left(\frac{\partial b}{\partial z}\right)^2 \right].    
    \end{split}
    \label{eq: Heating rate}
\end{equation}

\begin{figure*}[ht]
    \includegraphics[width=.33\textwidth]{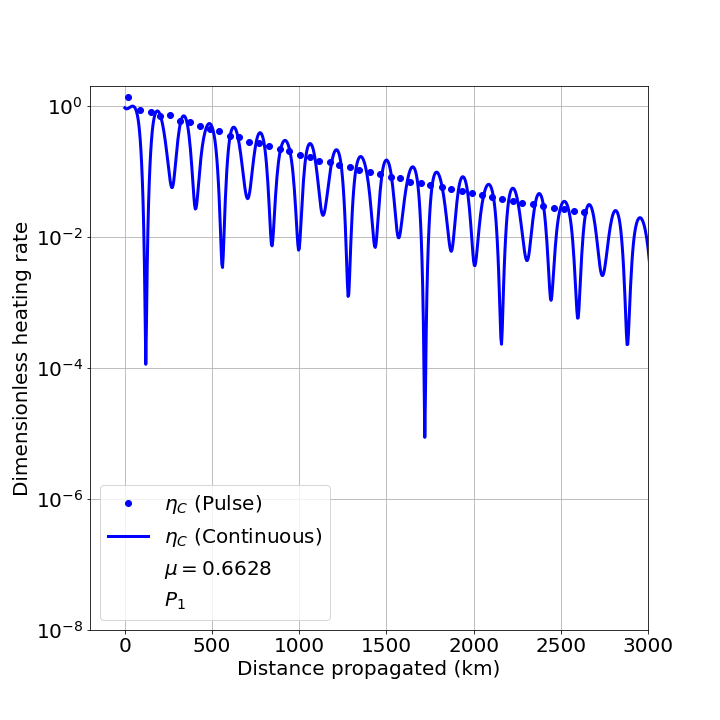}
    \includegraphics[width=.33\textwidth]{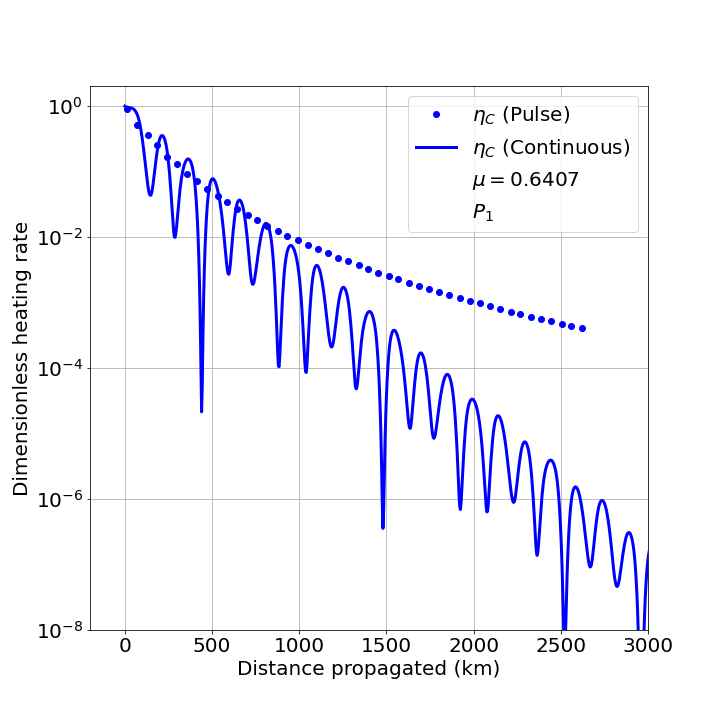}
    \includegraphics[width=.33\textwidth]{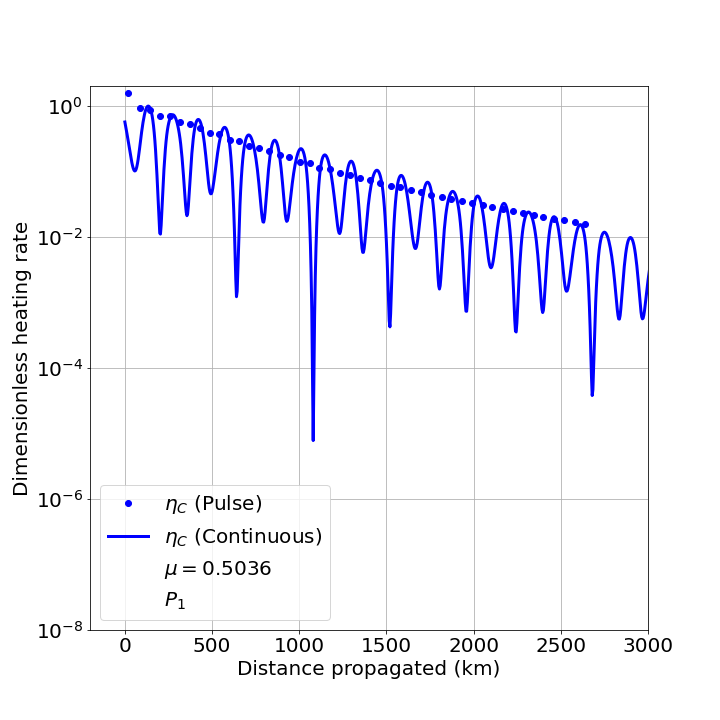}\hfill
    \\[\smallskipamount]
    \includegraphics[width=.33\textwidth]{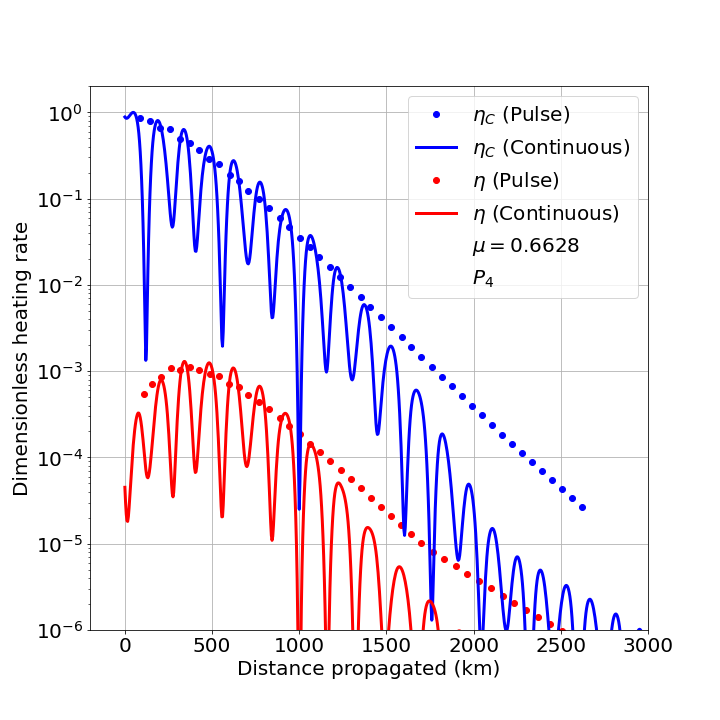}
    \includegraphics[width=.33\textwidth]{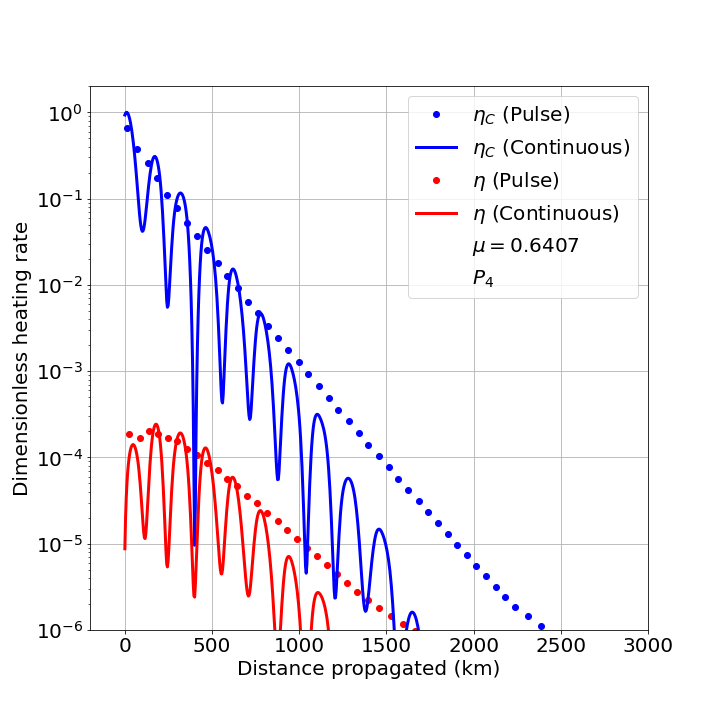}
    \includegraphics[width=.33\textwidth]{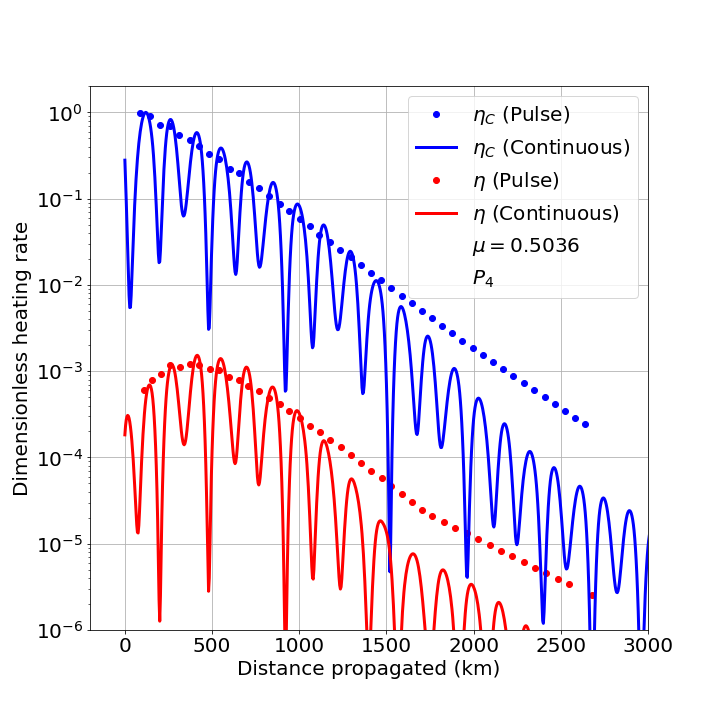}\hfill
    \caption{The profile of the heating rate obtained from the continuously excited sinusoidal wave driver (solid lines) is plotted with the tracked maximum heating rate of the Alfv\'en pulse (dotted line) over the propagated distance. The discrepancy between the results shown in the two rows arises from the existence of an inhomogeneous Alfv\'en speed profile, depicted in the right-hand column. Each column corresponds to a distinct ionization degree, consistent with the ionization degrees used in Figure \ref{fig:Continuous and finite} for the $P_1$ (upper row) and $P_4$ Alfv\'en speed wave profiles.}
    \label{fig:5.9}
\end{figure*}

Note that here we have omitted the viscous heating term from Equation (\ref{eq: Heating rate}), since it was found to play an insignificant role when compared with the two resistive heating terms in both the presently discussed simulations and those in \cite{McMurdo1}. The evolution of the heating rate of Alfv\'en pulses and continuously excited Alfv\'en waves are given as a function of distance propagated (shown in Figure \ref{fig:5.9}). 
Each heating term in Equation (\ref{eq: Heating rate}) depends on two quantities; the square of the derivative (transverse or longitudinal) and the dissipative coefficient (magnetic or Cowling diffusivity). Both derivatives are dependent on the amplitude of the wave and hence initially (at least for time steps up to the point where the pulse driver is switched off), the heating rates for the Alfv\'en pulse and the continuously driven Alfv\'en wave are identical since the pulses are generated by a continuous driver lasting only a finite time. 

In the case of the heating rate associated with longitudinal gradients, the heating rate decreases with propagation since the wavefront does not steepen and the wave is always decaying. The component of the heating related to cross-field derivatives increases with propagation and obtains its maximum once phase mixing has had time to develop. In this case, there are two competing aspects; the cross field derivative increases with propagation, however, the derivative is dependent on the amplitude of the wave which decreases with propagation and hence achieves a maximum some distance away from the wave source.

As was discussed by \cite{McMurdo1}, the maximum heating rate was found at the base of the domain, and for ionization degrees in the range $\mu = 0.5181 - 0.6570$ phase-mixed Alfv\'en waves provided sufficient heating to balance the radiative losses in the chromosphere (also confirmed observationally by \cite{Jess2009}). Since the maximal heating rate occurs near the wave source, the same conclusions can be drawn for the Alfv\'en pulses which means that phase-mixed Alfv\'en pulses can balance the radiative losses of the quiet solar chromosphere given an amplitude of $2.5$ km s$^{-1}$ propagating in a plasma with ionization degrees $\mu = 0.5181 - 0.6570$. Despite this obvious similarity, there is a considerable difference in the variation of the heating rate profile of Alfv\'en pulses compared with the heating rate profiles of Alfv\'en waves excited with a continuous driver.


The difference between the variation of the heating rates in the two cases starts to differ quite substantially as soon as the driver is turned off. The location of maximal heating rate propagates concurrently with the pulse and can be tracked and plotted against the heating profile obtained in the case of the continuously driven wave. The results of our analysis shown in Figure \ref{fig:5.9} are displayed for a range of ionization degrees and Alfv\'en speed profiles displayed in each panel. Note that the cases utilizing a homogeneous Alfv\'en speed profile, result in the absence of transversal gradients and hence the absence of a heating component corresponding to Ohmic diffusion. Consequently, only heating resulting from longitudinal gradients is observed.

Figure \ref{fig:5.9} illustrates the dimensionless heating rates derived from simulations involving both continuously driven waves and pulses. The highest combined heating rates occur at the base of the domain, primarily due to the magnitude of the Cowling diffusion, which is a few orders of magnitude larger than Ohmic diffusion. The Ohmic diffusion heating term increases with propagation until it peaks, then decreases as the perturbation amplitude diminishes, becoming the dominant factor in the heating rate. The dimensionless heating rate was chosen due to the dimensional quantity being highly sensitive on the chosen amplitude of the pulse (which may be taken from observations). 


As depicted in Figure \ref{fig:Continuous and finite}, instances arise where the amplitude of the pulses surpasses that of continuously driven waves, illustrated by the envelope plotted over the pulse profiles at various simulation time steps. The heating rate, given by Equation (\ref{eq: Heating rate}), is proportional to the square of the amplitude of the perturbation and the transversal and parallel gradients of the magnetic field perturbation. One of the effects that phase mixing has on the Alfv\'en pulses, not evident in the case of the continuously driven waves, is the widening of the pulse profile, causing the longitudinal gradients to reduce, subsequently leading to a lower maximum heating rate. These two competing effects of the increased amplitude yet reduced longitudinal gradients, result in the variation in heating profiles with propagation, as evidenced in Figure \ref{fig:5.9}. Moreover, while transversal gradients in continuously driven waves can only grow to a certain extent before naturally realigning, another limiting behavior restricts transverse gradients in Alfv\'en pulses. As the Alfv\'en pulses propagate, interactions between neighboring field lines result in elongation and flattening of the disturbances profile. This occurs due to the lack of a continuous energy injection at the base of the domain that preserves the wavelength of the disturbance with propagation, and results in the magnetic tension not allowing these transversal gradients to grow excessively large. This limitation constrains the component of the heating rate corresponding to Ohmic diffusion such that (for early times and distances propagated $\lesssim 500$ km) it does not exceed values observed in simulations of continuously driven Alfv\'en waves, contrary to initial expectations. Only later in the simulations did the heating rates corresponding to the Alfv\'en pulse significantly exceed those of the continually driven wave. This characteristic primarily stems from the greater amplitude of these disturbances at larger altitudes. While the heating rates obtained for pulses tend to exceed those excited by a continuous driver (at larger distances at least), this is attributed to the reduced damping rather than any particular efficiency in heating the solar atmosphere. The Alfv\'en pulses are, therefore, expected to propagate further into the solar atmosphere, carrying more energy to the corona, while continuously driven waves will dissipate more of their energy in the chromosphere. 


\section{Discussions}\label{sec:Discussion+Conclusions}

Our investigation sheds light on the intricate behavior of wave damping when considering finitely driven pulses. Our results, as depicted in Figure \ref{fig:Continuous and finite}, highlight instances where the amplitude of pulses surpasses that of continuously driven waves, leading to an impact on the evolution of the heating rate with propagated distance. We find that the heating rate, which depends on the square of both the perturbation amplitude and the gradients of the magnetic field perturbation, is influenced by the interaction between amplitude and the longitudinal and transverse gradients. Phase mixing of Alfv\'en pulses causes a broadening of the pulses profile, reducing longitudinal gradients and thereby lowering the heating rate associated with Cowling diffusion. This results in significant variations in heating profiles during propagation, as shown in Figure \ref{fig:5.9}. Furthermore, as displayed in Figure \ref{fig:Area change for vAs}, a significant enhancement is observed in the damping of phase-mixed pulses in the presence of inhomogeneous Alfv\'en speed profiles, particularly notable for the extremes of the simulated ionization degrees. Figures \ref{fig:Area change for wavelength}--\ref{fig:5.8} illustrate the variation in damping rates attributed to the frequency of the driver, aligning with conclusions drawn in \cite{McMurdo1}, where higher frequency waves display more effective damping.

While transverse gradients in the case of continuously driven waves exhibit a natural limit, Alfv\'en pulses show similar behavior, thus constraining their growth. As waves propagate, interactions between neighboring field lines elongate and flatten disturbances' profiles. This limitation on transverse gradients places a natural limit on the component of the heating rate corresponding to Ohmic diffusion. Interestingly, our simulations demonstrate that while heating rates corresponding to Alfv\'en pulses initially replicate those of continuously driven waves, they can significantly exceed them at later stages, for certain plasma configurations. This phenomenon primarily arises from the greater amplitude of finitely driven disturbances at higher altitudes. While waves with substantial amplitudes typically result in significant heating rates, the effectiveness of wave damping remains crucial. In general, we find the pulses damp to a lesser extent than the continually driven waves for the same propagated distance, suggesting that impulsively driven Alfv\'en waves have the potential to persist to longer distances and may play a more important role in heating the corona, compared with continuously driven waves. Our findings underscore the importance of understanding the complex interplay between the wave driver used in modeling, damping mechanisms, and propagation dynamics in resolving the atmospheric heating problem.

With respect to an Alfv\'en pulse's ability to heat the partially ionized plasma, a comparative study was made between impulsively driven and continuously driven waves. We conclude that phase-mixed pulses have the ability to balance radiative losses of the chromospheric region in the quiet sun, with the added characteristic, that they have the capability to traverse further distances into the solar atmosphere before undergoing full attenuation. We have presented evidence highlighting a distinction between the competing wave-driven mechanisms that may dominate heating in the partially ionized chromosphere and transmitting energy to the fully ionized corona. Specifically, impulsively driven waves appear to allow Alfv\'en waves to persist to higher altitudes where they may contribute to the heating of the corona, while continuously driven waves offer a more likely candidate to heat the chromosphere.

Before concluding, we wish to draw attention to some limitations of our model. Key ingredients were neglected to allow for a clear isolation of the effects of phase mixing and ionization degree, namely height and time-dependence of the dissipative coefficients and Alfv\'en speed profile as well as gravitational stratification. We expect that in a stratified environment, the damping rates of these Alfv\'en pulses will possess a rich form, where the local damping rate will vary depending on the local ionization degree and time dependent Alfv\'en speed profile.

\section{Acknowledgments}\label{sec:Acknowledgments}

MM received financial support from the Flemish Government under the long-term structural Methusalem funding program, project SOUL: Stellar evolution in full glory, grant METH/24/012 at KU Leuven. VF, GV and IB are grateful to the Science and Technology Facilities Council (STFC) grants ST/V000977/1, ST/Y001532/1. VF, GV thank the Institute for Space-Earth Environmental Research (ISEE, International Joint Research Program, Nagoya University, Japan). VF, GV, IB thank the Royal Society, International Exchanges Scheme, collaborations with Instituto de Astrofisica de Canarias, Spain (IES/R2/212183), Institute for Astronomy, Astrophysics, Space Applications and Remote Sensing, National Observatory of Athens, Greece (IES/R1/221095), Indian Institute of Astrophysics, India (IES/R1/211123) and collaboration with Ukraine (IES/R1/211177) for the support provided. This research was supported by the International Space Science Institute (ISSI) in Bern, through ISSI International Team project 457 (The Role of Partial Ionization in the Formation, Dynamics and Stability of Solar Prominences).

\bibliographystyle{aasjournal}

\bibliography{ref.bib}
\end{document}